\documentclass[12pt]{article}

\usepackage{amssymb,amsthm,amscd,array}
\usepackage{graphics}
\usepackage[final]{epsfig}

\let\ssection=\section
\renewcommand{\section}{\setcounter{equation}{0}\ssection}

\newcommand{\bbR}{\mathbb{R}}

\newcommand{\bbZ}{\mathbb{Z}}

\newcommand{\bbC}{\mathbb{C}}
\newcommand{\bbK}{\mathbb{K}}

\newcommand{\ad}{\mathrm{ad}}
\newcommand{\asl}{\mathrm{asl}}
\newcommand{\ah}{\mathrm{ah}}

\newcommand{\cD}{{\mathcal{D}}}
\newcommand{\cE}{{\mathcal{E}}}
\newcommand{\cI}{{\mathcal{I}}}

\newcommand{\cN}{{\mathcal{N}}}

\newcommand{\cK}{{\mathcal{K}}}
\newcommand{\cAK}{{\mathcal{AK}}}

\newcommand{\cP}{{\mathcal{P}}}

\newcommand{\E}{\mathrm{E}}
\newcommand{\cF}{{\mathcal{F}}}

\newcommand{\End}{\mathrm{End}}
\newcommand{\Der}{\mathrm{Der}}
\newcommand{\Id}{\mathrm{Id}}

\newcommand{\rk}{\mathrm{rk}}

\newcommand{\cJ}{\mathcal{J}}

\newcommand{\Sl}{\mathrm{sl}}
\newcommand{\gl}{\mathrm{gl}}

\newcommand{\aaf}{\mathrm{aaf}}
\newcommand{\osp}{\mathrm{osp}}
\newcommand{\OSp}{\mathrm{OSp}}

\newcommand{\half}{\frac{1}{2}}
\newcommand{\fg}{\mathfrak{g}}

\newcommand{\fa}{\mathfrak{a}}

\newcommand{\fC}{\mathfrak{C}}

\chardef\s=110
\chardef\g=103

\begin{document}

\font\rus=wncyr10 scaled 1000

\newtheorem{theorem}{Theorem}
\newtheorem{lemma}{Lemma}[section]
\newtheorem{cor}[lemma]{Corollary}
\newtheorem{conj}[lemma]{Conjecture}
\newtheorem{proposition}[lemma]{Proposition}
\newtheorem{rmk}[lemma]{Remark}
\newtheorem{exe}[lemma]{Example}
\newtheorem{defi}[lemma]{Definition}

\def\a{\alpha}
\def\b{\beta}
\def\d{\delta}
\def\e{\varepsilon}
\def\i{\iota}
\def\G{\Gamma}
\def\g{\gamma}
\def\L{\Lambda}
\def\om{\omega}
\def\r{\rho}
\def\s{\sigma}
\def\t{\tau}
\def\vfi{\varphi}
\def\vr{\varrho}
\def\l{\lambda}
\def\m{\mu}

\title{Lie antialgebras: \textit{pr\'emices}
}

\author{V. Ovsienko}

\date{}

\maketitle

\begin{abstract}
The main purpose of this work is to develop the
basic notions of the Lie theory for commutative algebras.
We introduce a class of $\bbZ_2$-graded commutative
but not associative algebras that we call
``Lie antialgebras''.
These algebras form a very special class of Jordan superalgebras,
they are closely related to Lie (super)algebras and,
in some sense, link together commutative and Lie algebras.
The main notions we define in this paper are:
representations of Lie antialgebras, an analog of the
Lie-Poisson bivector (which is not Poisson) and
central extensions.
We will explain the geometric origins of Lie antialgebras and
provide a number of examples.
We also classify simple finite-dimensional Lie antialgebras.
This paper is a new version of the unpublished preprint \cite{OvsLA}.
\end{abstract}

\thispagestyle{empty}

\tableofcontents

\section{Introduction}

Let $(\fa,\cdot)$ be a commutative $\bbZ_2$-graded algebra
over $\bbK=\bbR$ or $\bbC$,
that is,  $\fa=\fa_0\oplus\fa_1$, such that $\fa_i\cdot\fa_j\subset\fa_{i+j}$
and for all homogeneous elements $x,y$ one has
\begin{equation}
\label{SkewP}
x\cdot{}y\,=\,(-1)^{p(x)p(y)}
\,y\cdot{}x,
\end{equation}
where $p$ is the $\bbZ_2$-valued parity function
$p\vert_{\fa_i}=i$.
In particular, the space $\fa_0\subset\fa$ is a commutative subalgebra
while the bilinear map $\fa_1\times\fa_1\to\fa_0$ is skew-symmetric.

Typical examples of  $\bbZ_2$-graded commutative algebras are the
algebras of differential forms on manifolds, or,
more generally, the algebras of functions on supermanifolds;
these algebras are of course associative and, in particular, the space
$\fa_1$ is an $\fa_0$-module.
Examples of commutative but not associative algebras are
Jordan algebras.

The algebras considered in this paper are \textit{not associative},
however the subalgebra $\fa_0$ will always be associative.
In this sense, we suggest an alternative way to extend the
associativity condition of $\fa_0$ to all of the $\fa$.
We will try to convince the reader this leads to algebras that
have quite remarkable properties.

\subsection{The definition}

A $\bbZ_2$-graded commutative algebra $\fa$
is called a  \textit{Lie antialgebra} if it
satisfies the following identities:
\begin{eqnarray}
\label{AssCommT}
\a\cdot\left(\b\cdot{}\g\right)
&=&
\left(\a\cdot{}\b\right)\cdot{}\g,\\[10pt]
\label{CacT}
\a\cdot
\left(\b\cdot{}a\right)
&=&
\textstyle
\half
\left(\a\cdot{}\b\right)
\cdot{}a,\\[10pt]
\label{ICommT}
\a\cdot\left(a\cdot{}b\right)
&=&
\left(\a\cdot{}a\right)\cdot{}b
\;+\;
a\cdot{}
\left(\a\cdot{}b
\right),\\[10pt]
\label{Jack}
a\cdot\left(b\cdot{}c\right)
&+&
b\cdot\left(c\cdot{}a\right)
\;+\;
c\cdot\left(a\cdot{}b\right)
=0
\end{eqnarray}
where $\a,\b,\g\in\fa_0$ and $a,b,c\in\fa_1$.
In particular, $\fa_0$ is a (commutative) associative subalgebra of $\fa$.

\paragraph*
{Weaker form of (\ref{CacT})}
The identity (\ref{CacT}) needs a special discussion.
This identity means that for every $\a\in\fa_0$ the operator
$2\,\ad_\a:a\mapsto2\,\a\cdot{}a$  defines an
\textit{action} of the commutative algebra $\fa_0$ on the space $\fa_1$.
Furthermore, the identity (\ref{CacT}) implies
\begin{eqnarray}
\label{CacTbis}
\left(\a\cdot{}\b\right)
\cdot{}a=
\a\cdot
\left(\b\cdot{}a\right)+\b\cdot
\left(\a\cdot{}a\right)
\end{eqnarray}
which is a weaker identity.
The identity (\ref{CacTbis})
together with an additional assumption that the operators of multiplication by
even elements commute with each other,  is equivalent to (\ref{CacT}).
It turns out that the identity (\ref{CacTbis}) has an independent
algebraic meaning.

\paragraph*
{Axioms of Lie antialgebra and derivations}
Recall that $D\in\End(\fa)$ is a derivation of $\fa$ if
for homogeneous $x,y\in\fa$ (i.e., $x,y$ are either purely even or purely odd)
one has
\begin{equation}
\label{InvGrad}
D
\left(
x\cdot{}y
\right)=
D\left(x\right)\cdot{}y
+
(-1)^{p(D)p(x)}\,
x\cdot{}D\left(y\right).
\end{equation}
This formula then extends by linearity for arbitrary $x,y\in\fa$.
The space of all derivations of $\fa$ is a Lie superalgebra
denoted by $\mathrm{Der}(\fa)$.

Let us associate to every \textit{odd} element $a\in{}\fa_1$ the operator
$T_a:\fa\to{}\fa$ of \textit{right} multiplication by $a$
\begin{equation}
\label{RightRight}
T_a\left(x\right)=x\cdot{}a.
\end{equation}
The following observation
partly clarifies the definition of Lie antialgebras.

\bigskip

\noindent
{\it
The set of three identities: (\ref{CacTbis}),
(\ref{ICommT})  and (\ref{Jack})
is equivalent to the condition that  for all $a\in\fa_1$
the operator $T_a$ is a derivation.
}
\bigskip

\noindent

\paragraph*
{Associativity of $\fa_0$}
The associativity axiom (\ref{AssCommT}) seems quite different from the other
three axioms of Lie antialgebras.
For instance, it has no interpretation in terms of derivations.
It turns out however, that this axiom can be understood as a
corollary of the axioms (\ref{CacT})--(\ref{Jack}).

Assume that every even element of $\fa$ is a linear combination of
products of odd elements: $\a=\sum{}a_i\cdot{}a_j$.
We will call such a Lie antialgebra \textit{ample}.
The following simple statement is obtained in \cite{LMG}.

\bigskip

\noindent
{\it
If a Lie antialgebra $\fa$ is ample,
then the identities (\ref{CacT}), (\ref{ICommT}) 
and (\ref{Jack}) imply (\ref{AssCommT}).}

\bigskip

\noindent
Note that a similar property holds for Lie superalgebras.

\begin{rmk}
{\rm
It was noticed in \cite{LMG} that Lie antialgebras are closely related to
the Kaplansky superalgebras, see \cite{McC}.
More precisely, a half-unital Lie antialgebra is a Kaplansky superalgebra.
}
\end{rmk}

\subsection{Examples}

Let us give here a few examples of simple Lie antialgebras.

\bigskip

\textbf{1}.
Our first example is a $3$-dimensional Lie antialgebra called the 
\textit{tiny Kaplansky superalgebra}\footnote{This algebra was rediscovered in \cite{OvsLA}
where it was denoted by $\asl(2,\bbK)$, see also \cite{MG}.}
and denoted by $K_3$.
This algebra has the basis $\{\e;a,b\}$,
where $\e$ is even and $a,b$ are odd,
satisfying the relations
\begin{equation}
\label{aslA}
\begin{array}{l}
\e\cdot{}\e=\e\\[6pt]
\e\cdot{}a=\half\,a,
\quad
\e\cdot{}b=\half\,b,\\[6pt]
a\cdot{}b=\half\,\e.
\end{array}
\end{equation}
The algebra $K_3$ is simple, i.e., it contains no non-trivial ideal.
The corresponding algebra of derivations is the simple Lie superalgebra
$\osp(1|2)$, see \cite{BE} and  Section \ref{FirstAslS}.
Moreover, this property completely characterizes
the algebra $K_3$.

\begin{theorem}
\label{AsLThm}
The algebra $K_3$ is the unique finite-dimensional
$\bbZ_2$-graded commutative algebra such that the corresponding algebra of
derivations is isomorphic to $\osp(1|2)$.
\end{theorem}

\noindent
This theorem will be proved in Section \ref{UnSSe}.

\bigskip

\textbf{2}.
The most interesting example of a Lie antialgebra we know
is a simple infinite-dimensional Lie antialgebra with the basis
$
\textstyle
\left\{
\e_n,\;n\in\bbZ;
\quad
a_i,\;i\in\bbZ+\half
\right\},
$
where $\e_n$ are even and $a_i$ are odd
and satisfy the following relations
\begin{equation}
\label{GhosRel}
\begin{array}{rcl}
\e_n\cdot{}\e_m&=&
\e_{n+m},\\[8pt]
\e_n\cdot{}a_i &=&
\half\,a_{n+i},\\[8pt]
a_i\cdot{}a_j &=&
\left(j-i\right)\e_{i+j}.
\end{array}
\end{equation}
This algebra is called the 
{\it full derivation algebra}\footnote{It was called in \cite{OvsLA} the ``conformal Lie antialgebra''.},
see \cite{McC}. We will denote it by $\cAK(1)$.
We will prove that $\cAK(1)$ is closely related
to the well-known Neveu-Schwarz conformal Lie
superalgebra $\cK(1)$ namely
$$
\cK(1)=\mathrm{Der}(\cAK(1)).
$$
We conjecture that similarly to $K_3$
the algebra $\cAK(1)$ is the unique $\bbZ_2$-graded
commutative algebra satisfying this property.

The Lie antialgebra $\cAK(1)$ contains infinitely many copies of $K_3$
with the basis $\{\e_0;\,a_i,a_{-i}\}$ for each half-integer $i$.

\bigskip

\textbf{3}.
The subalgebra of $\cAK(1)$
with the basis 
$$
\left\{
\e_0,\,\e_1,\,\e_2,\ldots;
\;
a_{-\half},\,a_{\half},\,a_{\frac{3}{2}},\ldots
\right\}
$$
is also of some interest.
This algebra is simple and can be understood as
analog of the Lie algebra of (polynomial) vector fields on $\bbR$.
This algebra has interesting non-trivial cohomology
studied in \cite{LO}.

\bigskip

\textbf{4}.
Let us finally explain how to construct a large class of examples
of Lie antialgebras.
Given an arbitrary commutative algebra $\fC$,
there always exists an ample Lie antialgebra~$\fa$ such that $\fa_0=\fC\,$.
An example is provided by
$
K_3(\fC)=\fC\otimes_\bbK{}K_3(\bbK).
$
This already shows that there are at least as many ample Lie
antialgebras as there are commutative algebras.
However, this is not the only possibility
to realize a commutative algebra as an even par of a Lie antialgebra,
so that there are much more Lie antialgebras than commutative algebras.

\bigskip

We will give more concrete examples of Lie antialgebras in Section \ref{ClRS}.

\subsection{The main properties}

\paragraph*
{Lie antialgebras and Lie superalgebras and their representations}
Let $V$ be a $\bbZ_2$-graded vector space.
Consider the structure of Jordan $\bbZ_2$-graded algebra on
$\End(V)$ equipped with the anticommutator 
$[A,B]_+=AB+(-1)^{p(a)p(b)}\,BA$.
A \textit{representation} of a Lie superalgebra is a homomorphism
$$
\chi:\fa\to\left(\End(V),\,[.,.]_+\right)
$$
such that the \textit{image $\chi(\fa_0)$ is commutative},
i.e., $\chi(\a)\,\chi(\b)=\chi(\b)\,\chi(\a)$ for all $\a,\b\in\fa_0$.
This differs the notion of representation from the
well-known notion of specialization of a Jordan superalgebra.
Representations of $K_3$ were studied in \cite{MG},
study of representations of $\cAK(1)$ is an interesting problem,
see Section \ref{AntiCont} and \ref{RepS} for some examples.

The most interesting  property of Lie antialgebras is their relation to
Lie superalgebras.
With each Lie antialgebra $\fa$, we associate a Lie superalgebra $\fg_\fa$ 
in the following way.
The odd part $(\fg_\fa)_1$ coincides with $\fa_1$ while
the even part $\fg_0$ is the symmetric
tensor square $S^2(\fa_1)_{\fa_0}$,
where the tensor product is defined over the commutative algebra $\fa_0$.
Note that this construction is completely different from the
classical Kantor-Koecher-Tits construction.

\begin{itemize}

\item
Every representation of a Lie antialgebra $\fa$ is a representation of the
corresponding Lie superalgebra $\fg_\fa$.

\item
The Lie superalgebra $\fg_\fa$ acts on $\fa$, in other words, there is a
canonical homomorphism $\fg_\fa\to\Der(\fa)$.

\end{itemize}

\noindent
One can say that the Lie antialgebra $\fa$ selects a class of
representations of $\fg_\fa$ that are also representations of $\fa$.
This is an interesting characteristic of representations of
the Lie superalgebra $\fg_\fa$.

The properties of the Lie superalgebra $\fg_\fa$
and the universal enveloping algebras $U(\fg_\fa)$ and $U(\fa)$ 
will be studied with more details in \cite{LMG}.

\paragraph*
{ Odd ``Lie-Poisson'' bivector}
The notion of (odd) Lie-Poisson type bivector is the origin
of Lie antialgebras, see Section \ref{FirstAslS}.
For an arbitrary Lie antialgebra $\fa$,
the dual space with \textit{inverse parity,}~$\Pi\,\fa^*$,
can be equipped with a canonical odd linear bivector field $\L_\fa$,
see Section \ref{LiePoisson}.
Amazingly, the construction makes sense in the case of (purely even)
commutative associative algebra, i.e., for $\fa=\fa_0$, but the dual space should
be understood as purely odd in this case.
The bivector $\L_\fa$ is \textit{not Poisson}
in any sense.
Its general geometric characteristics is an interesting problem.

\paragraph*
{ Central extensions and cohomology}
In Section \ref{ExtS}, we introduce the notion of central extension of a Lie
antialgebra.
We prove
that existence of the unit element $\e\in\fa_0$
implies that the Lie antialgebra $\fa$ has no non-trivial central
extension.
Central extensions is a part of the general cohomology theory
developed in \cite{LO}.

\paragraph*
{ Classification}
In the finite-dimensional case, the classification
of simple Lie antialgebras is similar to
the classification of commutative division
algebras, see Section \ref{UnSSe}.
In the infinite-dimensional case, the situation is
of course much more complicated.
We also classify the Lie antialgebras of rank 1.

\section{Lie antialgebras and symplectic geometry}\label{First}

In this section, we show the way Lie antialgebras appear in geometry.
Notice that the invariant operations we construct are odd;
we recover Lie antialgebra structures using the parity inversion functor.

\subsection{The algebra $K_3$ and the odd bivector $\L$}\label{FirstAslS}

Consider the vector space $\bbK^{2|1}$ equipped with
the standard symplectic form, see \cite{Kos},
\begin{equation}
\label{SySt}
\textstyle
\om=dp\wedge{}dq+\half\,d\t\wedge{}d\t,
\end{equation}
where $p$ and $q$ are the usual even coordinates
on $\bbK^2$ and $\t$ is the formal Grassmann variable
so that $\t^2=0$.
An equivalent way to define this symplectic structure is to
introduce the Poisson bivector on
$\bbK^{2|1}$:
\begin{equation}
\label{PoiBiv}
\cP=\frac{\partial}{\partial{}p}
\wedge\frac{\partial}{\partial{}q}+
\half\,
\frac{\partial}{\partial{}\t}
\wedge\frac{\partial}{\partial{}\t}.
\end{equation}
which is inverse to the symplectic form: $\cP=\om^{-1}$.

The Lie superalgebra $\osp(1|2)$ is defined as the space
of infinitesimal linear transformations 
preserving the symplectic structure.
The bivector (\ref{PoiBiv}) is the unique
(up to a multiplicative constant) even
bivector invariant with
respect to the action of $\osp(1|2)$.

\paragraph*{The odd bivector}
It turns out that there exists another, \textit{odd},
$\osp(1|2)$-invariant bivector on $\bbK^{2|1}$.

\begin{proposition}
\label{IntLam}
There exists a unique
(up to a multiplicative constant)
odd bivector invariant with
respect to the action of $\osp(1|2)$.
It is give by the formula
\begin{equation}
\label{LaMb}
\Lambda=
\frac{\partial}{\partial\t}\wedge\cE+
\t\,\frac{\partial}{\partial{}p}
\wedge\frac{\partial}{\partial{}q},
\end{equation}
where
\begin{equation}
\label{EufD}
\cE=p\,\frac{\partial}{\partial{}p}+
q\,\frac{\partial}{\partial{}q}+
\t\,\frac{\partial}{\partial\t}
\end{equation}
is the Euler field.
\end{proposition}

\begin{proof}
The $\osp(1|2)$-invariance of $\Lambda$ is a very easy check.
Note that we will prove a much stronger statement,
see Proposition \ref{InvBivProp}.

Let us prove the uniqueness.
An arbitrary odd bivector on $\bbK^{2|1}$ is given by
$$
\Lambda=
\frac{\partial}{\partial\t}\wedge{}A+
\t\,F\,\frac{\partial}{\partial{}p}
\wedge\frac{\partial}{\partial{}q},
$$
where $A$ is an even vector field and $F$ is an even function.
Let $X$ be an even vector field, one has
$$
L_X\L=
\frac{\partial}{\partial\t}\wedge{}[X,A]+
\t\,X(F)\,\frac{\partial}{\partial{}p}
\wedge\frac{\partial}{\partial{}q}+
\t\,F\,L_X\left(
\frac{\partial}{\partial{}p}
\wedge\frac{\partial}{\partial{}q}
\right).
$$
If, furthermore, $X\in\osp(1|2)$, then it preserves the
even part $\frac{\partial}{\partial{}p}
\wedge\frac{\partial}{\partial{}q}$ of the Poisson  bivector.
The condition $L_X\L=0$ then implies:
$[X,A]=0$ and $X(F)=0.$

The even part of $\osp(1|2)$ is a Lie algebra isomorphic to
$\Sl(2,\bbK)$ and generated by the Hamiltonian vector fields with
quadratic Hamiltonians
$\langle{}p^2,\,pq,\,q^2\rangle$.
One checks that: 

a)~an even vector field $A$
commuting with any even element of
$\osp(1|2)$ is of the form
$$
A=c_1\t\,\frac{\partial}{\partial\t}+
c_2\,E,
$$
where $c_1$ and $c_2$ are arbitrary constants and
$
E=p\,\frac{\partial}{\partial{}p}
+q\,\frac{\partial}{\partial{}q};
$

b) an even function $F$ such that $X(F)=0$, where $X$
is an even element of
$\osp(1|2)$ is necessary a constant: $F=c_3$.

The odd part of $\osp(1|2)$ is spanned by the following two vector
fields:
$$
X_{\t{}p}=\t\,\frac{\partial}{\partial{}q}
+p\,\frac{\partial}{\partial\t},
\qquad
X_{\t{}q}=-\t\,\frac{\partial}{\partial{}p}
+q\,\frac{\partial}{\partial\t}.
$$
Applying any of these elements of $\osp(1|2)$ to the bivector
$\L$ as above, one
immediately gets $c_1=c_2=c_3$.
\end{proof}

The relation of the bivector $\L$ to the algebra
$K_3$ is as follows.
Any bivector defines an algebraic
structure on the space of functions.
Consider the bilinear operation associated
with the bivector (\ref{LaMb}):
\begin{equation}
\label{AntiPo}
\left]F,G\right[:=
\frac{(-1)^{p(F)}}{2}\,
\langle
\Lambda,dF\wedge{}dG
\rangle,
\end{equation}
where $F$ and $G$ are arbitrary functions on $\bbK^{2|1}$,
that is, $F=F_0(p,q)+\t\,F_1(p,q)$.
This is of course not a Poisson bracket.

\begin{lemma}
\label{SLeM}
The space of linear functions on $\bbK^{2|1}$ equipped
with the bracket (\ref{AntiPo}) is a Lie antialgebra
isomorphic to $K_3$.
\end{lemma}

\begin{proof}
One checks that after the \textit{parity inverting} identification
$\textstyle
\{\e;\,a,\,b\}
\longleftrightarrow
\{\t;\,p,\,q\},$
the algebra (\ref{aslA}) coincide with the bracket (\ref{AntiPo})
restricted to linear functions.
\end{proof}

We proved that the Lie superalgebra $\osp(1|2)$ preserves the bivector
$\L$.
Since $\osp(1|2)$ acts on $\bbK^{2|1}$ by linear vector fields,
it also preserves the space of linear functions.
In other words, $\osp(1|2)$ is the algebra of derivation
$\osp(1|2)=
\mathrm{Der}
\left(K_3)\right)$,
cf. \cite{BE}.

\begin{rmk}
{\rm
The bivector $\L$ given by (\ref{LaMb}),
and the Lie antialgebra $K_3$ are equivalent structures,
they contain the same information. 
The above lemma provides a parity inverting identification of the dual
space:
$$
\Pi\,K_3^*\cong(\bbK^{2|1},\L).
$$
The bivector $\L$ is therefore analog of the
``Lie-Poisson structure'' corresponding to $K_3$,
cf. Section \ref{LiePoisson} for a general setting.
}
\end{rmk}

\paragraph*{An algebraic reformulation}
A purely algebraic way to reformulate the results of this section
is as follows.

Consider the space of polynomials
$\bbK[p,q,\t]$ equipped with the standard action of
the Lie supergroup $\OSp(1|2)$
(or, equivalently, of the Lie superalgebra $\osp(1|2)$).
We are looking for $\OSp(1|2)$-invariant bilinear maps
$
(.,.):\bbK[p,q,\t]\otimes\bbK[p,q,\t]\to\bbK[p,q,\t]
$
satisfying the Leibniz rule in the both arguments, viz
$$
\left(FG,H\right)=
F\left(G,H\right)+(-1)^{p(G)p(H)}\left(F,H\right)G
$$
and similarly in the second argument.

This problem has exactly two solutions.

\begin{enumerate}
\item
The first operation is even, this is nothing but the standard Poisson bracket.
It can be defined at order one by
$$
\left\{p,q\right\}=1,
\qquad
\left\{p,\t\right\}=0,
\qquad
\left\{q,\t\right\}=0,
\qquad
\left\{\t,\t\right\}=1,
$$
and then extended to $\bbK[p,q,\t]$ via the Leibniz rule.
Polynoms of order 1 span the Heisenberg Lie superalgebra.

\item
The second operation is odd, it is defined at order 1 by
$$
\textstyle
\left]p,q\right[=\half\,\t,
\qquad
\left]p,\t\right[=\half\,p,
\qquad
\left]q,\t\right[=\half\,q,
\qquad
\left]\t,\t\right[=\t
$$
and, again, extends to $\bbK[p,q,\t]$ via the Leibniz rule.
This is the ``antibracket'' (\ref{AntiPo}), note that the
homogeneous polynomials of order 1 form an algebra
isomorphic to~$K_3$.
\end{enumerate}

\subsection{The full derivation algebra
$\cAK(1)$}\label{Conformal}

In this section we study the full derivation algebra $\cAK(1)$
defined by formula (\ref{GhosRel}).
We show that $\cAK(1)$ is simple and that
$\Der(\cAK(1))$ is isomorphic to the
famous conformal Lie superalgebra $\cK(1)$, also known as the
(centerless) Neveu-Schwarz algebra.

We also prove that the conformal Lie superalgebra $\cK(1)$
is the maximal algebra of vector fields on
$\bbR^{2|1}$ that preserves the bivector (\ref{LaMb}).
The algebra $\cAK(1)$ can be viewed as the maximal
space of functions on $\bbR^{2|1}$ that form a Lie antialgebra with
respect to the bracket (\ref{AntiPo}).

\paragraph*{The algebra $\cAK(1)$ is simple}
We start with the following

\begin{proposition}
\label{ConfAAProp}
The relations (\ref{GhosRel}) define a
structure of a simple Lie antialgebra.
\end{proposition}

\begin{proof}
The identities
(\ref{AssCommT}) and (\ref{CacT}) are evident.
Let us prove the identity (\ref{ICommT}).
One has to check that
$$
\e_n\cdot{}\left(a_i\cdot{}a_j\right)=
\left(\e_n\cdot{}a_i\right)\cdot{}a_j+
a_i\cdot{}\left(\e_n\cdot{}a_j\right)
$$
One obtains in the left-hand-side
$\half\left(j-i\right)\e_{i+j+n}$ and in the right-hand-side the sum of
two terms:
$\frac{1}{4}\left(j-(i+n)\right)\e_{i+j+n}$ and
$\frac{1}{4}\left(j+n-i\right)\e_{i+j+n}$, so that the identity
(\ref{ICommT}) is satisfied.
Furthermore, the identity (\ref{Jack}) reads:
$$
\left(a_i\cdot{}a_j\right)\cdot{}a_k+
\left(a_j\cdot{}a_k\right)\cdot{}a_i+
\left(a_k\cdot{}a_i\right)\cdot{}a_j=0.
$$
One obtains the sum $\frac{1}{4}\left(j-i\right)a_{i+j+n}+
\frac{1}{4}\left(k-j\right)a_{i+j+n}+
\frac{1}{4}\left(i-k\right)a_{i+j+n}=0$.
We proved that $\cAK(1)$ is, indeed, a Lie antialgebra.

It is quite easy to prove that $\cAK(1)$ is simple, see, e.g. \cite{McC}.
We do not dwell on the details here.
\end{proof}

\paragraph*{The conformal Lie superalgebra $\cK(1)$
as the algebra of derivations}
The conformal Lie superalgebra $\cK(1)$ has
the basis
$
\textstyle
\left\{
x_n,\;n\in\bbZ;
\quad
\xi_i,\;i\in\bbZ+\half
\right\}
$
satisfying the following commutation relations
\begin{equation}
\label{CAlgRel}
\begin{array}{rcl}
\left[
x_n,x_m
\right] &=&
\left(m-n\right)x_{n+m},\\[8pt]
\left[
x_n,\xi_i
\right] &=&
\left(i-\frac{n}{2}\right)\xi_{i+n},\\[8pt]
\left[
\xi_i,\xi_j
\right] &=&
x_{i+j}.
\end{array}
\end{equation}
It contains infinitely many copies of $\osp(1|2)$ with the
generators $\{x_{-n},x_0,x_n;\xi_{-\frac{n}{2}},\xi_\frac{n}{2}\}$.

Define the following action of $\cK(1)$ on
$\cAK(1)$:
\begin{equation}
\label{CAactRel}
\begin{array}{rcl}
x_n(a_i) &=&
\left(i-\frac{n}{2}\right)a_{n+i},\\[8pt]
x_n(\e_m) &=&
m\,\e_{n+m},\\[8pt]
\xi_i(a_j) &=&
\left(j-i\right)\e_{i+j},\\[8pt]
\xi_i(\e_n) &=&
a_{i+n}.
\end{array}
\end{equation}
Note that his formula is well-known and represents the action of the
 algebra $\cK(1)$ on the space of
tensor densities of weight $-\half$, see, e.g., \cite{Con,DM,GMO}
and Section~\ref{RepS}.

\begin{proposition}
\label{PaRo}
The action (\ref{CAactRel}) preserves the
structure of $\cAK(1)$.
\end{proposition}
\begin{proof}
Consider for instance the action of an odd element of $\cK(1)$.
One has
$$
\textstyle
\xi_i\left(a_j\cdot{}\e_k\right)=
\half\,\xi_i(a_{j+k})=
\frac{1}{2}\,(j+k-i)\,\e_{i+j+k},
$$
together with
$$
\textstyle
\xi_i(a_j)\cdot{}\e_k=
(j-i)\,\e_{i+j}\cdot{}\e_k=
(j-i)\,\e_{i+j+k}
$$
and
$$
\textstyle
a_j\cdot{}\xi_i(\e_k)=
a_j\cdot{}a_{i+k}=
\frac{1}{2}\,(j-i-k)\,\e_{i+j+k}.
$$
One finally gets:
$$
\xi_i\left(a_j\cdot{}\e_k\right)=
\xi_i(a_j)\cdot{}\e_k
-\,a_j\cdot{}\xi_i(\e_k)
$$
which is precisely the condition of odd derivation,
see formula (\ref{InvGrad}).

The action of other elements can be checked in the same way.
Hence the result.
\end{proof}

We will prove in the end of this section that $\cK(1)$ actually coincides with
$\Der(\cAK(1))$.

\paragraph*{Lie antialgebra $\cAK(1)$ and symplectic geometry}

In this section we show that, similarly to $K_3$,
the Lie antialgebra $\cAK(1)$ can be obtained from
the odd Poisson bivector~(\ref{LaMb}).

Consider the bracket (\ref{AntiPo}) given by the explicit expression
\begin{equation}
\label{AntibrExpF}
\left]F,G\right[=
\frac{(-1)^{p(F)}}{2}\left(
\frac{\partial{}F}{\partial\t}\,\cE(G)-
(-1)^{p(F)}\,\cE(F)\,\frac{\partial{}G}{\partial\t}+
\t\left(
\frac{\partial{}F}{\partial{}p}\,
\frac{\partial{}G}{\partial{}q}-
\frac{\partial{}F}{\partial{}q}\,
\frac{\partial{}G}{\partial{}p}
\right)
\right).
\end{equation}
One checks that the full antialgebra of functions
$C^\infty(\bbR^{2|1})$ equipped with this bracket
is \textit{not} a Lie antialgebra.

Let $\cF_\l$ be the space of
\textit{homogeneous functions of degree $-2\l$} on
$\bbR^{2|1}$, that is, of the functions
satisfying the condition
$$
\cE(F)=-2\l\,F,
$$
where $\cE$ is the Euler field (\ref{EufD}).
We will allow $F$ to have singularities in codimension 1,
for instance, we can consider rational functions.

Note that there is a dense subspace of homogeneous functions on $\bbR^{2|1}$
that correspond to the space of functions on $\bbR^{1|1}$.
Indeed, given a function
$
f(x,\xi)=f_0(x)+\xi\,f_1(x)
$
in one even and one Grassmann variable, one defines
a homogeneous of degree $\l$ function (with singularities at $(p=0)$) by
\begin{equation}
\label{Podem}
\textstyle
F^\l_f(p,q;\t)=
p^\l\,f
\left(\frac{q}{p},\,\frac{\t}{p}\right).
\end{equation}

\begin{proposition}
\label{HomAntilie}
The space, $\cF_{-\frac{1}{2}}$, of homogeneous of degree 1 functions on
$\bbR^{2|1}$ is a Lie antialgebra with
respect to the antibracket (\ref{AntibrExpF}) that contains
$\cAK(1)$.
\end{proposition}

\begin{proof}
The space of homogeneous of degree 1 functions is obviously stable
with respect to the antibracket (\ref{AntibrExpF}) so that $\cF_{-\frac{1}{2}}$ is an algebra.
Define a bilinear operation $f\cdot{}g$ on the space of functions in $(x,\xi)$ by
\begin{equation}
\label{AntibrLam}
F^1_{f\cdot{}g}:=
\frac{(-1)^{p(f)}}{2}
\left\langle
\L,dF^1_f\wedge{}dF^1_g
\right\rangle.
\end{equation}
One then easily checks the Lie antialgebra conditions.

Choose the following ``Taylor basis'':
$$
\textstyle
a_i=p\,\left(\frac{q}{p}\right)^{i+\half},
\qquad
\e_n=\t\left(\frac{q}{p}\right)^n
$$
and substitute it into the antibracket (\ref{AntiPo}).
One obtains the commutation relations (\ref{GhosRel}),
so that the Lie antialgebra $\cAK(1)$ is a subalgebra of $\cF_{-\frac{1}{2}}$.
\end{proof}

The conformal Lie superalgebra $\cK(1)$ also has a symplectic
realization.

\begin{proposition}
\label{HomConf}
The space $\cF_{-1}$ of homogeneous of degree 2 functions on
$\bbR^{2|1}$ is a Lie superalgebra with respect
to the Poisson bracket (\ref{PoiBiv}) that contains $\cK(1)$.
\end{proposition}

\begin{proof}
The Poisson bracket of two homogeneous of degree 2 functions
is, again, a homogeneous of degree 2 function.
Therefore, $\cF_{-1}$ is, indeed, a Lie superalgebra.

A homogeneous of degree 2 function can be written in
the form (\ref{Podem}) with $\l=2$.
Choosing the basis of the space of all such functions:
$$
\textstyle
x_n=\frac{p^2}{2}\,\left(\frac{q}{p}\right)^{n+1},
\qquad
\xi_i=\t{}p\left(\frac{q}{p}\right)^{i+\half}
$$
and substituting it into the Poisson bracket (\ref{PoiBiv}),
one immediately obtains the commutation relations (\ref{CAlgRel}).
Therefore, $\cK(1)$ is a subalgebra of $\cF_{-1}$.
\end{proof}

\begin{rmk}
{\rm
(a)
The Lie superalgebra $\cF_{-1}$ is a ``geometric version'' of the
conformal Lie superalgebra $\cK(1)$, which is a polynomial part of $\cF_{-1}$.
Similarly, $\cAK(1)$ is the polynomial part of the Lie antialgebra
$\cF_{-\half}$.

(b)
The action (\ref{CAactRel}) written in terms
of homogeneous functions
is, again, given by the standard Poisson bracket (\ref{PoiBiv}).
}
\end{rmk}

\paragraph*{Lie superalgebra $\cF_{-1}$ as algebra of symmetry}

\begin{proposition}
\label{InvBivProp}
The Lie superalgebra $\cF_{-1}$ is the maximal Lie superalgebra of
vector fields that preserves the
bivector (\ref{LaMb}).
\end{proposition}

\begin{proof}
\textit{Part 1}.
Let us first show that $\cF_{-1}$ preserves the bivector (\ref{LaMb}).
Given a function $H\in\cF_{-1}$, the corresponding
Hamiltonian vector field is homogeneous of degree 0:
\begin{equation}
\label{HHvf}
\left[
\cE,X_H
\right]=0.
\end{equation}

Consider a more general case, where
$P$ is a purely even (independent of $\t$) Poisson bivector
and $E$ a purely even vector field.
Assume $P$ is homogeneous
of degree $-2$ with respect to $E$, that is
$$
L_E(P)=-2\,P.
$$
Let $\L$ be the odd bivector field
$$
\L=
\frac{\partial}{\partial\t}\wedge\cE+
\t\wedge{}P,
$$
where
$$
\cE=E+\t\,\frac{\partial}{\partial\t}.
$$
(Note that in our case
$
E=
p\,\frac{\partial}{\partial{}p}+
q\,\frac{\partial}{\partial{}q},
\qquad
P=
\frac{\partial}{\partial{}p}
\wedge\frac{\partial}{\partial{}q}.
$)

Let $X_H$ be a Hamiltonian (with respect to the Poisson structure $P$)
vector field satisfying the homogeneity condition (\ref{HHvf}).
The Lie derivative of $\L$ along $X_H$ is as follows:
$$
\textstyle
L_{X_H}\L=
L_{X_H}\left(\frac{\partial}{\partial\t}\right)
\wedge\cE+
X_H\left(\t\right)P+
\t\,L_{X_H}\left(P\right).
$$
If $H$ is even, the above expression obviously vanishes.
Consider now an odd function $H=\t{}H_1$, then
one gets from (\ref{PoiBiv})
$$
X_{\t{}H_1}=
\t{}X_{H_1}+
H_1\,\frac{\partial}{\partial\t}.
$$

\begin{lemma}
\label{WeL}
One has
\begin{equation}
\label{CoBe}
L_{X_{\t{}H_1}}\left(\Lambda\right)=
\left\langle
P\wedge{}E,dH_1
\right\rangle
\end{equation}
where $d$ is the de Rham differential.
\end{lemma}

\begin{proof}
Using the obvious expressions
$$
\textstyle
\left[
X_{\t{}H_1},\frac{\partial}{\partial\t}
\right]
=
X_{H_1},
\qquad
\textstyle
L_{X_{\t{}H_1}}\left(P\right)=
-\frac{\partial}{\partial\t}\wedge{}X_{H_1},
$$
one obtains:
$$
\textstyle
L_{X_{\t{}H_1}}\left(\Lambda\right)=
X_{H_1}\wedge\cE+
H_1\,P+
\t\frac{\partial}{\partial\t}\wedge{}X_{H_1}=
X_{H_1}\wedge{}E+H_1\,P.
$$
Finally, using the fact that $E(H_1)=H_1$,
one obtains the expression (\ref{CoBe}).
\end{proof}

The even tri-vector $P\wedge{}E$ obviously vanishes
on $\bbR^{2|1}$, and so we proved that $X_H$, indeed, preserves the
bivector (\ref{LaMb}).

\textit{Part 2}.
Conversely, one has to show that any vector field preserving the
bivector (\ref{LaMb}) is a Hamiltonian vector field commuting with
$\cE$.

If $X$ is a purely even vector field, i.e.,
$$
\textstyle
\left[
X,\frac{\partial}{\partial\t}
\right]=0
\quad
\hbox{and}
\quad
X(\t)=0,
$$
then $L_X(\L)=0$ implies that $X$ commutes with $\cE$ and preserves the
even bivector
$P=\frac{\partial}{\partial{}p}
\wedge\frac{\partial}{\partial{}q}$,
so that $X$ is Hamiltonian.

If $X$ is an odd vector field:
$$
X=F_0\,\frac{\partial}{\partial\t}+\t\,X_0,
$$
where $F_0$ is an even function and $X_0$ an even vector field, then
one obtains explicitly
$$
L_{X}\L=
\left(
E(F_0)-F_0
\right)
\frac{\partial}{\partial\t}
\wedge\frac{\partial}{\partial\t}
-
\t\frac{\partial}{\partial\t}
\wedge
\left(
X_{F_0}+[E,X_0]
\right)
+
\left(
F_0\,P+
X_0\wedge{}E
\right).
$$
The assumption $L_X(\L)=0$ implies that each of the three summands in
this expression vanishes.
It follows from the equation
$
E(F_0)-F_0=0,
$
that $F_0$ is a homogeneous of degree 1
function.
The condition
$$
X_{F_0}+[E,X_0]=0
$$
then implies that $X_0$ is a vector
field homogeneous of degree $-1$, since so is $X_{F_0}$, and thus
$X_0=X_{F_0}$.
We proved that the vector field $X$ is Hamiltonian and $[\cE,X]=0$.

Proposition \ref{InvBivProp} is proved.
\end{proof}

\begin{cor}
\label{DerCor}
One has
$$
\cF_{-1}=\Der(\cF_{-\half}),
\qquad
\cK(1)=\Der(\cAK(1)).
$$
\end{cor}

\noindent
Indeed, the first statement is a reformulation of Proposition \ref{HomConf}
while the second is its algebraic version.
The subalgebra $\cK(1)\subset\cF_{-1}$ corresponds precisely to the
space of vector fields preserving the basis of $\cAK(1)\subset\cF_{-\half}$.

\subsection{Representation
of $\cAK(1)$ by tangent vector fields}\label{AntiCont}

In this section we investigate the relation of the  Lie antialgebra
$\cAK(1)$ to contact geometry.
In some sense, $\cAK(1)$ provides a way to
``integrate'' the contact structure.

\paragraph*{The contact structure on $S^{1|1}$}

The natural projection
$
\bbR^{2|1}\setminus\{0\}\longrightarrow{}S^{1|1},
$
equips $\bbR^{1|1}$ with a structure of $1|1$-dimensional contact
manifold, see \cite{Con,GMO,DM} for recent developments.
This contact structure can be defined in terms of a contact
1-form
$
\a=dx+\xi{}d\xi,
$
or, equivalently, in terms of an odd
vector field\footnote{This vector field is also known in physical
literature as ``SUSY-structure''.}
$$
D=\half\left(
\frac{\partial}{\partial\xi}
+\xi\frac{\partial}{\partial{}x}
\right),
$$
since $D$ spans the kernel of $\a$.

The conformal Lie superalgebra $\cK(1)$ can be realized as the Lie
superalgebra of contact vector fields on $S^{1|1}$.
Every contact vector field on $S^{1|1}$ is of the form
$$
X_h=
h(x,\xi)\,\frac{\partial}{\partial{}x}+
2\,D\left(h(x,\xi)\right)\,D,
$$
where $h(x,\xi)=h_0(x)+\xi\,h_1(x)$ is an arbitrary function.
The map $h\mapsto{}F^2_h$, see (\ref{Podem}), is an
isomorphism between Lie superalgebra of contact vector fields
and $\cF_{-1}$.
The Lie superalgebra of contact vector fields with polynomial coefficients
is isomorphic to $\cK(1)$.

\paragraph*{Vector fields tangent to the contact distribution}

It turns out that there is a similar relation between the
algebra $\cAK(1)$ and the contact geometry.

A vector field \textit{tangent} to the contact distribution is a vector
field proportional to $D$, that is, $X=f\,D$ for some function
$f(x,\xi)$.

\begin{defi}
{\rm
We introduce the following
\textit{anticommutator} of tangent vector fields:
\begin{equation}
\label{AntiLieBr}
\left[f\,D,g\,D\right]_+:=
f\,D\circ{}g\,D+
(-1)^{(p(f)+1)(p(g)+1)}\,g\,D\circ{}f\,D.
\end{equation}
Note that the sign in this operation is inverse to that of usual
commutator of vector fields.
}
\end{defi}

\begin{rmk}
{\rm
The space of tangent vector fields is not a Lie superalgebra since
the Lie bracket of two tangent vector fields is not a tangent vector
field, this is equivalent to non-integrability of the contact
distribution.
}
\end{rmk}

Quite miraculously, that the anticommutator of two tangent vector
fields is again a tangent vector
field.

\begin{proposition}
\label{TProp}
The space of tangent vector fields equipped with the
anticommutator (\ref{AntiLieBr}) is a Lie antialgebra
that contains the Lie antialgebra $\cAK(1)$.
More precisely,
$$
\left[\chi_f,\chi_g\right]_+=
\chi_{f\cdot{}g},
$$
where $f\cdot{}g$ is the product (\ref{AntiPo}).
\end{proposition}

\begin{proof}
Define a map from $\cAK(1)$ to the space of tangent vector fields
as follows.
To each homogeneous of degree 1 function $F^1_f(p,q,\t)$, cf. formula
(\ref{Podem}), we associate a tangent vector field by
\begin{equation}
\label{FRep}
\chi_f=
f(x,\xi)\,D.
\end{equation}
Let us calculate the explicit formula of the anticommutator
(\ref{AntiLieBr}).

Consider first the case where both of the functions, 
$f$ and $g$, are odd, i.e.,
$f=\xi{}f_1$ and $g=\xi{}g_1$.
One then has
$$
f\,D\circ{}g\,D+g\,D\circ{}f\,D=
\half\,fg\,\frac{\partial}{\partial{}x}+
\left(f\,D(g)+g\,D(f)\right)D.
$$
The first summand is zero since it contains $\xi^2=0$, while the
second one is equal to $\xi\left(f_1g_1\right)D$, so that the
anticommutator (\ref{AntiLieBr}) on the odd functions coincides with
$
f_1g_1.
$
This corresponds to the product $f_1\cdot{}g_1$ of two elements of the even
part $\cAK(1)_0$, see formula (\ref{GhosRel}).

If $f=\xi{}f_1$ is odd and $g=g_0$ is even, then the Leibniz rule
$D\circ{}f=D(f)-f\,D$ implies
$$
\textstyle
f\,D\circ{}g\,D+g\,D\circ{}f\,D=
g\,D(f)\,D=
\half\,f_1g_0\,D,
$$
so that one gets
$
f_1\cdot{}g_0=\half\,f_1g_0,
$
accordingly to the $\cAK(1)_0$-action on $\cAK(1)_1$, cf.
(\ref{GhosRel}).

If, finally, the both functions $f$ and $g$ are even,
i.e., $f=f_0$ and $g=g_0$, then
$$
\textstyle
f\,D\circ{}g\,D-g\,D\circ{}f\,D=
\left(f\,D(g)-g\,D(f)\right)D=
\half\,\xi\left(f_0g_0'-g_0f_0'\right)D,
$$
gives the skew-symmetric product
$
f_0\cdot{}g_0=
\half\left(
f_0g'_0-g_0f'_0
\right),
$
on $\cAK(1)_1$ with values in
$\cAK(1)_0$, see formula (\ref{GhosRel}).
\end{proof}

\begin{rmk}
{\rm
The map (\ref{FRep}) is nothing but the
bivector (\ref{LaMb}) contracted with the elements of $\cAK(1)$.
One checks that
\begin{equation}
\label{LepLam}
\chi_f(g)=
\frac{(-1)^{p(f)}}{2}
\left\langle
\L,dF^1_f\wedge{}dF^0_g
\right\rangle,
\end{equation}
where $F^1_f$ and $F^0_g$ are functions on $\bbR^{2|1}$
homogeneous of degree $1$ and $0$, respectively,
obtained as the lift of $f$ and $g$ according to (\ref{Podem}).
It is interesting to compare the above formula with
(\ref{AntibrLam}).
}
\end{rmk}

We conclude this section with the formula for the product on
the space of smooth functions on $S^{1|1}$ that \textit{coincides} with (\ref{GhosRel})
on the polynomial basis.
The functions are of the form
$f(x,\xi)=\a(x)+\xi\,a(x)$, the product is given by
\begin{equation}
\label{GhosRelSmooth}
\textstyle
\left(\a+\xi\,a\right)\cdot\left(\b+\xi\,b\right)=
\a\b+ab'-a'b\,+\,
\half\,\xi\left(
\a{}b+\b{}a
\right).
\end{equation}
This is the $C^\infty$-analog of $\cAK(1)$.

\subsection{A pair of symplectic forms on $\bbR^{4|2}$
and the algebras $K_3(\bbC)$ and $\cAK(1)_\bbC$}\label{SimRk2}

Consider the space $\bbK^{2n|m}$ with linear symplectic form.
There is no analog of the odd bivector (\ref{LaMb}), if $n\geq2$ or $m\geq2$,
since the only $\OSp(m|2n)$-invariant bivector is
the Poisson bivector.
In this section we investigate the second special case: $n=m=2$.
It turns out that one has to fix a \textit{pair} of symplectic forms
and consider the group of linear transformations preserving
both of them.

Note that, if $n\geq3$ or $m\geq3$, then one has to consider degenerate
linear 2-forms, cf. Section \ref{LiePoisson}.

\paragraph*{A pair of symplectic forms on $\bbK^{4|2}$}

Consider two \textit{generic} linear symplectic forms on $\bbK^{4|2}$.
There exist linear coordinates
$(p_1,q_1,p_2,q_2;\t_1,\t_2)$ such that these forms are as follows.
$$
\textstyle
\om_\e=
dp_1\wedge{}dq_1+
dp_2\wedge{}dq_2+
\half\left(
d\t_1\wedge{}d\t_1-
d\t_2\wedge{}d\t_2
\right)
$$
and 
$$
\om_\s=
dp_1\wedge{}dp_2-
dq_1\wedge{}dq_2+
d\t_1\wedge{}d\t_2.
$$
This is a very simple fact of linear algebra.

Note that, in the real case, the forms $\om_\e$ and $\om_\s$ are, respectively,
the real and the imaginary parts of the
complex 2-form (\ref{SySt}),
where $p,q;\t$ are the following complex coordinates
$
p=p_1+iq_2,
\,
q=q_1+ip_2,
\,
\t=\t_1+i\t_2.
$

\begin{proposition}
\label{RealComSy}
The Lie superalgebra of
linear vector fields on $\bbK^{4|2}$ preserving the two symplectic
forms $\om_\e$ and $\om_\s$ is isomorphic to:

(i)
$\osp(1|2,\bbC)$,
if $\bbK=\bbR$;

(ii)
$\osp(1|2,\bbC)\oplus\osp(1|2,\bbC)$,
if $\bbK=\bbC$;

\end{proposition}

\begin{proof}
The Lie superalgebra of
linear vector fields on $\bbK^{4|2}$ preserving the two symplectic
forms $\om_\e$ and $\om_\s$ is a subalgebra of $\osp(2|4,\bbK)$ spanned by
6 even Hamiltonian vector fields corresponding to the quadratic
Hamiltonians
$$
\left\{
p_1^2-q_2^2,\quad
p_2^2-q_1^2,\quad
p_1p_2+q_1q_2,\quad
p_1q_1-p_2q_2,\quad
p_1q_2,\quad
q_1p_2
\right\}
$$
and 4 odd bi-Hamiltonian vector fields with the Hamiltonians
$$
\left\{
p_1\t_1-q_2\t_2,\quad
p_2\t_1+q_1\t_2,\quad
q_1\t_1-p_2\t_2,\quad
q_2\t_1+p_1\t_2
\right\}.
$$
In the real case, this defines an $\osp(1|2,\bbC)$-action on $\bbR^{4|2}$.
In the complex case, this is $\osp(1|2,\bbC)_\bbC\cong\osp(1|2,\bbC)\oplus\osp(1|2,\bbC)$.
\end{proof}

\begin{rmk}
{\rm
Note that, in the case (i), $\osp(1|2,\bbC)$ is viewed as a simple Lie superalgebra
over $\bbR$.
This algebra of derivations is thus obtained as
intersection of two copies of the real symplectic Lie superalgebra:
$$
\osp(1|2,\bbC)=
\osp_\e(1,1\,|\,4)\cap\osp_\s(1,1\,|\,4)
$$
corresponding to the symplectic forms $\om_\e$ and $\om_\s$,
respectively.
}
\end{rmk}

\paragraph*{The bivector $\L^C$}

The following bivector is the unique (up to a multiplicative
constant) odd $\osp(1|2,\bbC)$-invariant bivector on $\bbK^{4|2}$:
\begin{equation}
\label{LaMbRk2}
\Lambda^C=
\frac{\partial}{\partial\t_1}\wedge\cE+
\frac{\partial}{\partial\t_2}\wedge{}\cJ+
\t_1\,\pi_\e+
\t_2\,\pi_\s,
\end{equation}
where 
$$
\cE=
\sum_{i=1,2}
\left(
p_i\,\frac{\partial}{\partial{}p_i}+
q_i\,\frac{\partial}{\partial{}q_i}+
\t_i\,\frac{\partial}{\partial{}\t_i}
\right)
$$
 is the Euler vector field and
\begin{equation}
\label{VFAsigma}
\cJ=
q_2\,\frac{\partial}{\partial{}p_1}+
p_2\,\frac{\partial}{\partial{}q_1}
-q_1\,\frac{\partial}{\partial{}p_2}
-p_1\,\frac{\partial}{\partial{}q_2}+
\t_2\,\frac{\partial}{\partial\t_1}-
\t_1\,\frac{\partial}{\partial\t_2}
\end{equation}
and where the bivectors $\pi_\e$ and $\pi_\s$ are
$$
\pi_\e=
\frac{\partial}{\partial{}p_1}
\wedge\frac{\partial}{\partial{}q_1}+
\frac{\partial}{\partial{}p_2}
\wedge\frac{\partial}{\partial{}q_2},
\qquad
\pi_\s=\frac{\partial}{\partial{}p_1}
\wedge\frac{\partial}{\partial{}p_2}-
\frac{\partial}{\partial{}q_1}
\wedge\frac{\partial}{\partial{}q_2}.
$$

It is now easy to check that
the space of linear functions
on $(\bbR^{4|2},\L^C)$ form a Lie antialgebra isomorphic to $K_3(\bbC)$.

\paragraph*{Algebra $\cAK(1)_\bbC$ and
the pair of symplectic structures on $\bbR^{4|2}$}

Let us now realize the algebras
$\cAK(1)_\bbC$ and $\cK(1)_\bbC$ in terms of real
rational harmonic functions on $\bbR^{4|2}$.
This is related to the bi-Hamiltonian formalism
defined by the pair of symplectic structures
$\om_\e$ and $\om_\s$.

The Lie antialgebra $\cAK(1)_\bbC$ is represented by the homogeneous
of degree 1 harmonic functions on $\bbR^{4|2}$ with the complex
structure (\ref{VFAsigma}).
The odd generators of $\cAK(1)_\bbC$ are as follows
$$
\textstyle
\e_n=
\frac{1}{2^n}\,\mathrm{Re}\,
\t
\left(\frac{q}{p}\right)^n,
\quad
\b_n=
\frac{1}{2^n}\,\mathrm{Im}\,
\t
\left(\frac{q}{p}\right)^n,
$$
and the even ones are
$$
\textstyle
a_n=
\frac{1}{2^n}\,\mathrm{Re}\,
p
\left(\frac{q}{p}\right)^n,
\quad
b_n=
\frac{1}{2^n}\,\mathrm{Im}\,
p
\left(\frac{q}{p}\right)^n,
$$
where $p,q,\t$ are the complex coordinates.
One then checks the relations in $\cAK(1)_\bbC$.

The conformal Lie superalgebra $\cK(1)_\bbC$ is realized by
homogeneous bi-Hamiltonian vector fields on $\bbR^{4|2}$.
For each function $F\in{}C^\infty(\bbR^{4|2})$, denote by
$X^\e_F$ and $X^\s_F$ the Hamiltonian vector fields on $\bbR^{4|2}$
with respect to the symplectic form $\om_\e$ and $\om_\s$,
respectively.

One checks that the following three conditions are
equivalent.

\begin{enumerate}
\item
$F$ is a homogeneous of degree 2 harmonic function:
$$
\textstyle
F=
\mathrm{Re}\,
p^2f
\left(\frac{q}{p},\frac{\t}{p}
\right),
$$
where $f$ is an arbitrary function.

\item
The function $F$ satisfies the relations:
$\cE(F)=2F$ and $\cJ\left(\cJ\left(F\right)\right)=-4F$.

\item
The Hamiltonian vector fields with the Hamiltonian $F$ commute with
$\E$ and $\cJ$:
$$
[\cE,X^\e_F]=[\cE,X^\s_F]=[\cJ,X^\e_F]=[\cJ,X^\s_F]=0
$$
and are bi-Hamiltonian such that
$$
\textstyle
X^\e_F=-\half\,X^\s_{\cJ(F)},
\qquad
X^\s_F=\half\,X^\e_{\cJ(F)}.
$$
\end{enumerate}

This space of homogeneous harmonic bi-Hamiltonian vector fields is a Lie
superalgebra isomorphic to $\cK(1)_\bbC$.

\begin{proposition}
\label{CompMaxPro}
The Lie superalgebra $\cK(1)_\bbC$ is the maximal algebra of
vector fields on $\bbR^{4|2}$ preserving the bivector (\ref{LaMbRk2}).
\end{proposition}

\begin{proof}
Proposition \ref{InvBivProp} implies that $\cK(1)_\bbC$, indeed, preserves
the bivector $\L^C$.
The proof of maximality is similar to that of Theorem \ref{InvBivProp},
Part 2.
We omit here the corresponding straightforward computations.
\end{proof}

\section{Elements of the general theory}\label{General}

In this section we investigate the relations between Lie antialgebras
and Lie superalgebras.
We introduce the notion of representation and module
over a Lie antialgebra and show how to extend them to
the corresponding Lie superalgebra.

We also define the analog of the Lie-Poisson structure for an arbitrary Lie
antialgebra~$\fa$.
This is an odd linear bivector field on the dual space with inverse parity $\Pi\,\fa^*$.
This notion relates Lie antialgebras to geometry.

We finally define and study the notion of central extensions
that will later be useful for classification results.

\subsection{Relation to Lie superalgebras}\label{Quad}

In this section we associate a Lie superalgebra $\fg_\fa$
with an arbitrary Lie antialgebra $\fa$.
The construction will be important since representations of a Lie
antialgebra can always be extended to the corresponding Lie superalgebra.

\begin{defi}
{\rm
Consider the space
$
\fg_\fa=
\left(\fg_\fa\right)_0\oplus
\left(\fg_\fa\right)_1$
defined as follows.
The even part, $\left(\fg_\fa\right)_0$, is the symmetric tensor square of
$\fa_1$ over $\fa_0$:
$$
\left(\fg_\fa\right)_0:=\fa_1\odot_{\fa_0}\fa_1,
$$
more precisely,
$$
\left(\fg_\fa\right)_0=
\left\{
(a\otimes{}b+b\otimes{}a)/_\sim
\;|\;
a,b\in\fa_1
\right\},
$$
where the equivalence relation $\sim$ is defined by 
$$
\a\cdot{}a\otimes{}b\sim{}a\otimes\a\cdot{}b
$$
for all $a,b\in\fa_1$ and $\a\in\fa_0$.
The odd part of $\left(\fg_\fa\right)$, is nothing but the odd part of $\fa$:
$$
\fg_1:=\fa_1.
$$
The Lie bracket on $\fg_\fa$ is defined by
\begin{equation}
\label{CommKvad}
\begin{array}{rcl}
\left[
a\odot{}b\,,\,c\odot{}d
\right]&=&
a\cdot{}\left(b\cdot{}c\right)\odot{}d
+b\cdot{}\left(a\cdot{}c\right)\odot{}d
-c\cdot{}\left(d\cdot{}a\right)\odot{}b
-d\cdot{}\left(c\cdot{}a\right)\odot{}b,\\[10pt]
\left[
a\odot{}b\,,\,c
\right]&=&
a\cdot{}\left(b\cdot{}c\right)+
b\cdot{}\left(a\cdot{}c\right),\\[10pt]
\left[
a\,,\,b
\right]&=&
a\odot{}b.
\end{array}
\end{equation}
}
\end{defi}

\begin{theorem}
\label{SqThm}
The space $\fg_\fa$ endowed with the bracket
(\ref{CommKvad}) is a Lie superalgebra.
\end{theorem}

\noindent
The Jacobi identity and the
compatibility with the equivalence $\sim$ for the bracket
(\ref{CommKvad}) can be checked by a
complicated but straightforward calculation that we omit.

\paragraph*{The action of $\fg_\fa$ on $\fa$}
Let us define a Lie superalgebra homomorphism
$
T:\fg_\fa\to\Der(\fa).
$
The restriction $T\left|_{\left(\fg_\fa\right)_1}\right.$ is naturally defined
since the odd part $\left(\fg_\fa\right)_1$ is identified with $\fa_1$ and
the odd part $\fa_1$ acts on $\fa$ by derivations, cf. formula (\ref{RightRight}).
One then extends this map to the even part $\left(\fg_\fa\right)_0$
in a unique way as follows:
$T_{a\odot{}b}:=[T_a,T_b]$.
One then has explicitly
\begin{equation}
\label{Qder}
T_{a\odot{}b}\,x=
\left(x\cdot{}a\right)\cdot{}b+
\left(x\cdot{}b\right)\cdot{}a
\end{equation}
for odd $a,b\in\fa_1$ and arbitrary $x\in\fa$.

\begin{lemma}
\label{SQprop}
(i)
The map (\ref{Qder}) is well-defined, that is,
it is compatible with the equivalence relation $\sim$.

(ii)
For all $a,b\in\fa_1$, the
linear map $T_{a\odot{}b}\in\End(\fa)$
is an even derivation.
\end{lemma}

\begin{proof}
Part (i).
The computation is straightforward but we provide here
some details in order to show how work the axioms of Lie antialgebra.

One has to check that
$$
\left(x\cdot{}(\a\cdot{}a)\right)\cdot{}b
+\left(x\cdot{}b\right)\cdot{}\left(\a\cdot{}a\right)
=
\left(x\cdot{}(\a\cdot{}b)\right)\cdot{}a
+\left(x\cdot{}a\right)\cdot{}\left(\a\cdot{}b\right)
$$
for every $\a\in\fa_0$ and arbitrary $x$.
Consider, for instance, the case where $x$ is odd, that is $x\in\fa_1$.
Using (\ref{CacT}), one obtains
$$
\left(x\cdot{}b\right)\cdot{}\left(\a\cdot{}a\right)
=
\a\cdot{}\left((x\cdot{}b)\cdot{}a\right)
$$
while using (\ref{ICommT}) and then (\ref{Jack}) and again (\ref{CacT}), one obtains 
$$
\begin{array}{rcl}
\left(x\cdot{}(\a\cdot{}a)\right)\cdot{}b
&=&
\left(\a\cdot{}(x\cdot{}a)\right)\cdot{}b
-\left((\a\cdot{}x)\cdot{}a\right)\cdot{}b
\\[6pt]
&=&
2\,\a\cdot{}\left((x\cdot{}a)\cdot{}b\right)
+\left(a\cdot{}b\right)\cdot{}\left(\a\cdot{}x\right)
+\left(b\cdot{}(\a\cdot{}x)\right)\cdot{}a
\end{array}
$$
for the terms in the left-hand-side.
Similarly, for the right-hand-side one has
$$
\begin{array}{rcl}
\left(x\cdot{}a\right)\cdot{}\left(\a\cdot{}b\right)
&=&
\a\cdot{}\left((x\cdot{}a)\cdot{}b\right)
\\[6pt]
\left(x\cdot{}(\a\cdot{}b)\right)\cdot{}a
&=&
2\,\a\cdot{}\left((x\cdot{}b)\cdot{}a\right)
-\left((\a\cdot{}x)\cdot{}b\right)\cdot{}a.
\end{array}
$$
Finally, collecting the terms in the $(\hbox{lhs})-(\hbox{rhs})$ one gets
$$
\begin{array}{rcl}
\a\cdot{}\left((x\cdot{}a)\cdot{}b\right)
-\a\cdot{}\left((x\cdot{}b)\cdot{}a\right)
+\left(a\cdot{}b\right)\cdot{}\left(\a\cdot{}x\right)
&=&
\\[6pt]
-\a\cdot\left((a\cdot{}b)\cdot{}x\right)
+\left(a\cdot{}b\right)\cdot{}\left(\a\cdot{}x\right)
&=&0,
\end{array}
$$
using (\ref{Jack}).

The computation for an even element $x\in\fa_0$ is similar.

Part (ii) is obvious since
$T_{a\otimes{}b}$ is a commutator of two derivations.
\end{proof}

\begin{rmk}
{\rm
The above proof makes an extensive use of the axioms
(\ref{CacT})--(\ref{Jack}) of Lie antialgebra.
The associativity axiom (\ref{AssCommT}) is
never used, but, as already mentioned, this axiom is
also a corollary of (\ref{CacT})--(\ref{Jack}).
The complete proof of Theorem \ref{SqThm} is quite
similar but much longer and more complicated, see \cite{LMG}.
}
\end{rmk}

\begin{exe}
{\rm
In the case of the simple algebras
$\fa=K_3$ and $\cAK(1)$,
the corresponding Lie superalgebras $\fg_\fa$
coincide with the respective algebra of derivations: 
$$
\fg_{K_3}=\osp(1|2),
\qquad
\fg_{\cAK(1)}=\cK(1).
$$
However, in general this is not the case.
The more Lie antialgebra $\fa$ is far of being simple, the more the
corresponding Lie superalgebra $\gl(\fa)$ is far of $\Der(\fa)$.
}
\end{exe}

\subsection{Representations and modules}\label{RepS}

In this section we introduce important notions
of representation and of module over Lie antialgebras.
Amazingly, these two notions are different for Lie antialgebras.
We will investigate these notions in some details for the
algebras $K_3(\bbC)$ and $\cAK(1)$.

\paragraph*{Representations: the definition}
Let $V=V_0\oplus{}V_1$ be a $\bbZ_2$-graded vector space.
Consider the Jordan algebra structure on the space $\End(V)$
defined by the anticommutator
\begin{equation}
\label{AntiCoCo}
\left[A,B\right]_+=
A\,B+(-1)^{p(A)p(B)}
B\,A.
\end{equation}
Note that this operation the sign rule inverse to that of the usual
$\bbZ_2$-graded commutator
(the sign is ``$-$'' if and only if both $A$ and $B$ are odd).

\begin{rmk}
{\rm
Note that the Jordan algebra $\left(\End(V),\,[.,.]_+\right)$ is
of course not a Lie antialgebra.
However, we use this algebra to define the notion of representation.
}
\end{rmk}

Given a Lie antialgebra $\fa$, an (even) linear map
$\chi:\fa\to\End(V)$
is called a \textit{representation} of $\fa$, if two conditions
are satisfied:
\begin{equation}
\label{TrueR}
\chi_{x\cdot{}y}=\left[\chi_x,\chi_y\right]_+,
\end{equation}
for all $x,y\in\fa$ and the images of even elements are commuting operators:
\begin{equation}
\label{TrueRCommm}
\chi_\a\,\chi_\b=\chi_\b\,\chi_\a
\end{equation}
for all $\a,\b\in\fa_0$.

\begin{rmk}
{\rm
A linear map $\chi:\fa\to\End(V)$ satisfying (\ref{TrueR}), but not necessarily
(\ref{TrueRCommm}) is called a {\it specialization} of a Jordan algebra.
This means, a representation is a very particular case of specialization.
The condition (\ref{TrueRCommm}) is crucial for us.
}
\end{rmk}

\begin{exe}
{\rm
We have already considered a particular case of the anticommutator
(\ref{AntiCoCo}), namely the operation (\ref{AntiLieBr})
on the space of vector fields on $S^{1|1}$ tangent to the contact structure.
The map (\ref{FRep}) defines a representation of the full derivation
algebra $\cAK(1)$ in the space of differential operators on
$\bbR^{1|1}$.
}
\end{exe}

\begin{theorem}
\label{ExtPr}
Every representation of a Lie antialgebra $\fa$ is
naturally a representation of the corresponding Lie superalgebra $\fg_\fa$.
\end{theorem}

\begin{proof}
Given a representation $\chi$ of a Lie antialgebra $\fa$,
the construction of the corresponding representation of $\fg_\fa$
is as follows.
The odd part $\left(\fg_\fa\right)_1$
coincides with $\fa_1$, so that the map $\chi\vert_{\left(\fg_\fa\right)_1}$
is already defined.
For the even elements of $\fg_\fa$ we take the usual commutator:
$$
\chi_{a\odot{}b}:=
\left[
\chi_a,\chi_b
\right]=
\chi_a\circ\chi_b+\chi_b\circ\chi_a.
$$
To prove Theorem \ref{ExtPr}, one has to show that this is
indeed a representation of $\fg_\fa$.

The complete proof is again a very complicated but straightforward computation, 
 that we omit.
The details are given in \cite{LMG}.
\end{proof}

The above result shows that representations of $\fa$ is some particular
class of representations of $\fg_\fa$.

\paragraph*
{Representations of $K_3$}
Representations of the simple algebra $K_3$
were studied in \cite{MG}.

The corresponding Lie superalgebra is $\osp(1|2)$, so that
representations of $K_3$ is a particular class of representations
of $\osp(1|2)$.
It turns out that this class is characterized in terms of the
classical Casimir element $C\in{}U(\osp(1|2))$.
The following statement is the main result of \cite{MG}.

\bigskip

{\it There is a one-to-one correspondence between
representations of s Lie antialgebra $K_3$ and the
representations of the Lie superalgebra $\osp(1|2)$ such that
$\chi(C)=0$.}

\bigskip

A representation of $K_3$ is obviously
given by one even operator $\cE\in\End(V)$ and two odd operators
$A,B\in\End(V)$ satisfying the relations
\begin{equation}
\label{TwistHeis}
\begin{array}{rcl}
AB-BA&=&\cE,\\[6pt]
A\cE+\cE{}A&=&A,\\[6pt]
B\cE+\cE{}B&=&B,\\[6pt]
\cE^2&=&\cE.
\end{array}
\end{equation}
Among other classification results of \cite{MG}, let us
mention the following simple but beautiful statement.

\bigskip

{\it Up to isomorphism, the operator $\cE$ in (\ref{TwistHeis}) is of the form:
\begin{equation}
\label{OpEr}
\cE\vert_{V_0}=0,
\qquad
\cE\vert_{V_1}=\Id.
\end{equation}}

\bigskip

\noindent
The relations (\ref{TwistHeis}) together with (\ref{OpEr}) are very similar
to the Heisenberg canonical commutation relations.
The difference is that the operator $\cE$ is not the identity but the
projector to the odd part of $V$.

\paragraph*{Example: tensor-density representations of $\cK(1)$ and $\cAK(1)$}

Consider the representations of $\cK(1)$
called tensor density representations $\cF_\l$,
where $\l\in\bbC$ is the parameter.
The basis in $\cF_\l$ is $\{f_n,\,n\in\bbZ,\quad\phi_i,\,i\in\bbZ+\half\}$
and the action of  $\cK(1)$ is given by
$$
\begin{array}{rcl}
\chi_{x_n}(f_m) &=&
\left(m+\l{}n\right)f_{n+m},\\[8pt]
\chi_{x_n}(\phi_i) &=&
\left(i+(\l+\half)n\right)\phi_{n+i},\\[8pt]
\chi_{\xi_i}(f_n) &=&
\left(\frac{n}{2}+\l{}i\right)\phi_{i+n},\\[8pt]
\chi_{\xi_i}(\phi_j) &=&
f_{i+j}.
\end{array}
$$

For instance, the adjoint representation of $\cK(1)$ is
isomorphic to $\cF_{-1}$.
The $\cK(1)$-action (\ref{CAactRel}) on $\cAK(1)$ is isomorphic to
$\cF_{-\half}$ with inverse parity.

It is now very easy to check that
the $\cK(1)$-module $\cF_\l$ is a representation of $\cAK(1)$
if and only if
$$
\l=0
\qquad
\hbox{or}
\qquad
\l=\half.
$$
Note that the modules $\cF_0$ and $\cF_\half$ are dual to each other,
these two modules are known to be special.

\paragraph*{Modules over Lie antialgebras: the definition}

A $\bbZ_2$-graded vector space $V$
is called an $\fa$-\textit{module} if
there is an even linear map
$
\rho:\fa\to\End(V),
$
such that the direct sum $\fa\oplus{}V$ equipped
with the product
\begin{equation}
\label{TrueM}
(a,\,v)\cdot(b,\,w)
=
\left(
a\cdot{}b\,,\,
\r_a\,w+(-1)^{p(b)p(v)}\,\r_b\,v
\right),
\end{equation}
where $a,b\in\fa$ and $v,w\in{}V$ are homogeneous elements,
is again a Lie antialgebra.
We call the Lie antialgebra structure (\ref{TrueM})
a \textit{semi-direct} product and denote it by
$\fa\ltimes{}V$.

\begin{exe}
{\rm
1).
The ``adjoint action''
$\ad:\fa\to\End(\fa)$ defined by
$$
\ad_x\,y=x\cdot{}y,
$$
for all $x,y\in\fa$, defines a structure of $\fa$-module on $\fa$ itself.
This follows, for instance, from the fact that the tensor product
$\fC\otimes\fa$ of a Lie antialgebra $\fa$ with a commutative algebra
$\fC$ is again a Lie antialgebra.
Indeed, consider $\fC=\bbK[t]/(t^2)$, then one has
$
\fC\otimes\fa=\fa\ltimes\fa.
$

2).
The ``coadjoint action''
$\ad:\fa\to\End(\fa^*)$ defined by
$$
\ad_x\,\vfi=(-1)^{p(x)p(\vfi)}\ad^*_x\vfi,
$$
for $x\in\fa$ and $\vfi\in\fa^*$, defines a structure of $\fa$-module on $\fa^*$.
}
\end{exe}

Note that the maps $\ad$ and $\ad^*$ are \textit{not} a representation of $\fa$ since
these maps do not satisfy (\ref{TrueR}).

\subsection{The odd Lie-Poisson type bivector}\label{LiePoisson}

In this section we introduce the notion of canonical
odd bivector field $\L_\fa$ on the dual space $\Pi\,\fa^*$.

\paragraph*{The rank and the pencil of presymplectic forms}

Given a Lie antialgebra $\fa$,
we call the dimension of the
odd part $\fa_0$ of a Lie antialgebra $\fa$ the \textit{rank} of
$\fa$:
$$
\rk\,\fa:=\dim\,\fa_0.
$$

Let $\fa$ be a Lie antialgebra of rank $r$.
Fix an arbitrary basis
$\{\e_1,\ldots,\e_r\}$ of $\fa_0$.
One obtains a set of $r$ bilinear
skew-symmetric (or presymplectic) forms: $\{\om_1,\ldots,\om_r\}$ on
$\fa_1$ by projection on the basic elements:
\begin{equation}
\label{OmFor}
a\cdot{}b=:
\sum_{i=1}^r\om_i(a,b)\,\e_i,
\end{equation}
for all $a,b\in\fa_1$.

Changing the basis, one obtains linear changes of the
corresponding set of skew-symmetric forms.
Therefore, the \textit{pencil of presymplectic forms}
$
\langle\om_1,\ldots,\om_r\rangle
$
is well-defined.
To each 2-form $\om_i$, one associates a
bivector
$
\pi_i\in\wedge{}^2\fa^*_1,
$
that we can understand as a constant bivector field on $\fa^*_1$.

\paragraph*{The adjoint vector fields}

To each element $\a\in\fa_0$, one associates an even linear
operator $A_\a:\fa\to\fa$ defined by
\begin{equation}
\label{AOp}
A_\a\left|_{\fa_0}\right.=\ad_\a,
\qquad
A_\a\left|_{\fa_1}\right.=2\,\ad_\a.
\end{equation}
These linear operators can be, of course,
viewed as \textit{linear vector fields} on $\fa^*$.

We will denote $A_1,\ldots,{}A_r$ the (even) vector fields on $\fa^*$
corresponding to the elements of the basis $\{\e_1,\ldots,\e_r\}$.

\begin{exe}
{\rm
In the important case where $\fa_0$ contains the unit element
$\e$ and the center of $\fa$ is trivial (see Theorem \ref{ZenT}
below), one has
$
A_\e=\cE,
$
where $\cE$ is the Euler vector field on the vector space $\fa^*$,
i.e., the generator of the $\bbK^*$-action by homotheties.
}
\end{exe}

\paragraph*{The definition}

We will use the
parity inversion functor $\Pi$.
Consider the space $\Pi\fa_0^*$
and denote by $(\t_1,\ldots,\t_r)$ the \textit{Grassmann} coordinates dual
to the chosen basis.
Choose furthermore arbitrary linear coordinates $(x_1,\ldots,x_n)$ on $\Pi\fa_1^*$.

Define the following odd bivector on $\Pi\fa^*$:
\begin{equation}
\label{PLBiv}
\Lambda_\fa=
\sum_{i=1}^r
\left(
\frac{\partial}{\partial\t_i}\wedge{}A_i+
\t_i\,\pi_i
\right).
\end{equation}
The corresponding antibracket on the space of (polynomial, smooth, etc.) functions on
$\Pi\fa^*$ is defined as in (\ref{AntiPo}).
This antibracket is obviously linear, i.e., the space of
linear functions on $\Pi\fa^*$ is stable.

\begin{proposition}
\label{LieProp}
Linear functions on the space $(\Pi\fa^*,\L_\fa)$ form a Lie antialgebra
isomorphic to $\fa$.
\end{proposition}

\begin{proof}
The antibracket of two even linear functions obviously corresponds to
(\ref{OmFor}).
The odd linear functions on $\Pi\fa^*$ are linear combinations of
$\t_1,\ldots,\t_r$.
The antibracket of an odd and an even linear functions
is given by
$
]\t_i,\ell[=\half\,A_i(\ell),
$
where $\ell\in\fa_0$.
This corresponds precisely to the adjoint action of $\e_i$ on $\ell$.

Finally, the antibracket of two odd linear functions is given by
$$
\textstyle
]\t_i,\t_j[=
\half\left(
A_i(\t_j)+A_j(\t_i)
\right)
\leftrightarrow
\half\left(
\ad_{\e_i}\e_j+\ad_{\e_j}\e_i
\right)=
\e_i\cdot\e_j.
$$
Hence the result.
\end{proof}

\begin{cor}
\label{NiceCor}
The bivector (\ref{PLBiv}) is independent of the choice of the basis.
\end{cor}

\begin{exe}
\label{EulEx}
{\rm
The bivectors (\ref{LaMb}) and (\ref{LaMbRk2})
are precisely the canonical bivectors on
$\Pi\,K_3(\bbR)^*$ and $\Pi\,K_3(\bbC)^*$, respectively.
}
\end{exe}

\begin{exe}
\label{CoBiv}
{\rm
The bivector (\ref{LaMb}) makes sense
in the case of a (purely even) commutative algebra,
i.e., where $\fa_1=\{0\}$.
Let $\fC$ be a commutative associative algebra with basis
$\{\e_1,\ldots,\e_r\}$ and structural constants $c_{ij}^k$ such
that $\e_i\cdot\e_j=\sum_kc_{ij}^k\,\e_k$.
Let $\t_1,\ldots,\t_r$ be a set of Grassmann coordinates.
The bivector (\ref{PLBiv}) is then of the form
$$
\L_\fC=
\sum_{i,j,k}\,c_{ij}^i\,\t_k\,
\frac{\partial}{\partial{}\t_i}\wedge\frac{\partial}{\partial{}\t_j}.
$$
We understand this linear bivector field as commutative analog of the
Lie-Poisson bivector.
}
\end{exe}

\paragraph*{The conformal Lie superalgebra}
The Lie superalgebra $\Der(\fa)$ can be viewed as the algebra
of \textit{linear} vector fields on $\Pi\fa^*$ preserving the bivector $\L_\fa$.
It is natural to define (an infinite-dimensional) Lie superalgebra of
``conformal derivations'' of $\fa$ as the algebra of
vector fields on $\Pi\fa^*$ preserving the bivector $\L_\fa$.
We simply drop the linearity condition.

\begin{exe}
{\rm
The algebra $\cK(1)$ is the algebra of conformal derivations of 
$K_3$.
}
\end{exe}

The notion of algebra of conformal derivations of $\fa$
deserves a special study.
It would be also interesting to understand if there is a general
notion of conformal Lie antialgebra.
It makes sense to look for such a notion in terms of
the algebra of functions on $\Pi\fa^*$ homogeneous
with respect to the vector fields
$A_1,\ldots,{}A_r$.

\subsection{Central extensions}\label{ExtS}

In this section we define the notion of extension
of a Lie antialgebra $\fa$ with coefficients in any
$\fa$-module.
It will be useful for the classification
result of Section \ref{Class}.
The notion of extension is a part of a
general cohomology theory that will be developed in \cite{LO}.

Let $\fa$ be a Lie antialgebra and $V$ an $\fa$-module.
We will consider $V$ as a trivial (or abelian) Lie
antialgebra.

\begin{defi}
\label{ExtDef}
{\rm
(a)
An exact sequence of Lie antialgebras
\begin{equation}
\label{SubmodDiagramm}
\begin{CD}
0 @> >>V @> >> \widetilde\fa @> >> \fa @> >>0
\end{CD}
\end{equation}
is called an \textit{abelian extension} of the Lie antialgebra $\fa$
with coefficients in $V$.
As a vector space, $\widetilde\fa=\fa\oplus{}V$, and
the subspace $V$ is obviously an $\fa$-module.

(b)
An extension (\ref{SubmodDiagramm}) is called \textit{non-trivial} if
the Lie antialgebra
$\widetilde\fa$ is not isomorphic to the semi-direct sum
$\fa\ltimes{}V$.

(c)
If the subspace $V$ belongs to the center of $\widetilde\fa$,
then the extension (\ref{SubmodDiagramm}) is called a
\textit{central extension}.
}
\end{defi}

In this section we develop the notion of central extension.
Since any central extension can be obtained by iteration of
one-dimensional central extensions, it suffice to consider only the
case
of one-dimensional central extensions.
One then has two possibilities:
$$
\dim{}V=0|1,
\quad
\hbox{or}
\quad
\dim{}V=1|0.
$$
We then say that the one-dimensional central extension is of
\textit{type I} or of \textit{type II}, respectively.

\paragraph*{Central extensions of type I}
The general form of central extensions of type I is as follows.

\begin{proposition}
\label{T1CeP}
(i)
A central extension of type I is defined by (an even) symmetric bilinear map
$C:\fa\otimes\fa\to{}\bbK^{0|1}$ satisfying the the following conditions:
\begin{equation}
\label{ExtI}
\begin{array}{rcl}
\displaystyle
C\left(\a,\,\b\cdot{}a\right)\,=\,
C\left(\b,\,\a\cdot{}a\right)&=&
\half\,C\left(\a\cdot{}\b,\,a\right)\\[6pt]
\displaystyle
C\left(a\cdot{}b,\,c\right)+
C\left(b\cdot{}c,\,a\right)+
C\left(c\cdot{}a,\,b\right)&=&0,
\end{array}
\end{equation}
for all $\a,\b\in\fa_0$ and $a,b,c\in\fa_1$.

(ii)
A central extension is trivial if and only if there exists
an even linear function $f:\fa_0\to\bbK$ such that
\begin{equation}
\label{Cob}
C(\a,a)=f(\a\cdot{}a)
\end{equation}
for all $\a\in\fa_1$ and $a\in\fa_0$.
\end{proposition}

\begin{proof}
Part (i).
Given a map $C$ as in (\ref{ExtI}), let us
define an antibracket on $\fa\oplus\bbK$.
We fix an element $z\in\bbK$ and set:
\begin{equation}
\label{NewStrI}
x\cdot{}y
=
(x\cdot{}y)_\fa+
C(x,y)\,z,
\qquad
x\cdot{}z
=0,
\end{equation}
for all $x,y\in\fa$.
Note that, since the map $C$ is symmetric and even, one has
\begin{equation}
\label{TakC}
C(x,y)=
C\left(x_1,y_0\right)+
C\left(y_1,x_0\right),
\end{equation}
where $x_1,y_1\in\fa_1$ and $x_0,y_0\in\fa_0$.
One then easily checks that formula (\ref{NewStrI}) defines a
structure of a Lie antialgebra if and only if the relations
(\ref{ExtI}) are satisfied.

Conversely, a Lie antialgebra structure on $\fa\oplus\bbK^{0|1}$ such that
the subspace $\bbK^{0|1}$ belongs to the center is obviously of the form
(\ref{NewStrI}).

Part (ii).
In the case where $C$ is as in (\ref{Cob}), the linear map
$\fa\oplus\bbK^{0|1}\to\fa\oplus\bbK^{0|1}$ given by $(x,z)\mapsto(x,z+f(x))$
intertwines the structure (\ref{NewStrI}) with the trivial direct sum
structure. This means that the central extension is trivial.

Conversely, if the extension is trivial, then there exists an
intertwining
map $\fa\oplus\bbK^{0|1}\to\fa\oplus\bbK^{0|1}$ sending the structure
(\ref{NewStrI})
to the trivial one.
This map can, again, be chosen in the form $(x,z)\mapsto(x,z+f(x))$,
since
a different choice of the embedding of $\bbK^{0|1}$ does not change the
structure.
\end{proof}

We will call a map $C$ satisfying (\ref{ExtI}) a 2-\textit{cocycle of
type} I. A 2-cocycle of the form (\ref{Cob}) will be called a
\textit{coboundary}.

\paragraph*{Central extensions of type II}

\begin{proposition}
\label{T2CeP}
(i)
A central extension of type II is defined by an even symmetric bilinear map
$
C:\fa\otimes\fa\to\bbK
$
satisfying the following identities:
\begin{equation}
\label{ExtII}
\begin{array}{rcl}
C\left(\a,\,a\cdot{}b\right)&=&
C\left(\a\cdot{}a,\,b\right)+
C\left(a,\,\a\cdot{}b\right)\\[8pt]
C\left(\a\cdot{}\b,\,\g\right)&=&
C\left(\a,\,\b\cdot{}\g\right),
\end{array}
\end{equation}
for all $\a,\b,\g\in\fa_0$ and $a,b\in\fa_1$.

(ii)
The extension is trivial if and only if
there exists an even linear functional $f:\fa_1\to\bbK$ such that
\begin{equation}
\label{CobII}
C(x,y)=f(x\cdot{}y),
\end{equation}
for all $x,y\in\fa$.
\end{proposition}

\begin{proof}
The proof is similar to that of Proposition \ref{T1CeP}.
\end{proof}

We will call an odd map (\ref{ExtII}) satisfying (\ref{ExtII}) a
2-\textit{cocycle of type} II. In the case where it is given by
(\ref{CobII}), the map $C$ is called a coboundary.

\medskip

\begin{exe}
{\rm
Consider the
kernel of the presymplectic pencil of $\fa$:
$$
\cI=\bigcap_{i=1}^r\ker\om_i,
$$
where $r=\rk\,\fa$, the forms $\om_i$ are defined by
(\ref{OmFor}).
In other words,
$$
\cI=\{a\in\fa_1\;|\;a\cdot{}b=0,\; \hbox{for all}\; b\in\fa_1\}.
$$

\begin{proposition}
\label{KerLem}
The subspace $\cI$ is an abelian ideal of $\fa$.
\end{proposition}
\begin{proof}
By definition, $\cI$ is an abelian subalgebra and the bracket
of $a\in\cI$ with any element $b\in\fa_1$ vanishes.
One has to show that $\a\cdot{}a\in\cI$,
for arbitrary $\a\in\fa_0$ and $a\in\cI$.
Indeed, using the identity (\ref{ICommT}), one obtains
$$
\left(
\a\cdot{}a
\right)\cdot{}b=
\a\cdot{}\left(
a\cdot{}b
\right)-
a\cdot{}\left(
\a\cdot{}b
\right)=0
$$
since for $a\in\cI$ and every $b\in\fa_1$, one has
$a\cdot{}b=0$.
\end{proof}

It follows that the Lie antialgebra $\fa$ itself is an abelian
extension of type I of the quotient-algebra $\fa/\cI$:
$$
\begin{CD}
0 @> >>I @> >> \fa @> >> \fa/\cI @> >>0.
\end{CD}
$$
Let us also outline the case where this extension has to be central.

\begin{proposition}
\label{ZenLem}
If $\fa$ is ample, then 
the ideal $\cI$ belongs to the center of $\fa$.
\end{proposition}

\begin{proof}
The ideal $\cI$ belongs to the center of $\fa$ if and only if
the action of $\fa_0$ on $\cI$ is trivial.
Recall that $\fa$ is ample if map $\fa_1\otimes\fa_1\to\fa_0$ is surjective.
Surjectivity means that for every $\a\in\fa_0$
there are $a,b\in\fa_1$ such that
$\a=a\cdot{}b$.
Using the Jacobi identity (\ref{Jack}), one obtains for every
$c\in\cI$:
$$
\a\cdot{}c
=
\left(a\cdot{}b\right)\cdot{}c
=
-
\left(b\cdot{}c\right)\cdot{}a-
\left(c\cdot{}a\right)\cdot{}b
=0,
$$
since both summands in the right-hand-side vanish.
\end{proof}

The Lie antialgebra $\fa$ is therefore a central extension of $\fa/\cI$
(of type I).
}
\end{exe}

\paragraph*{The case of the unit element}
Let us consider the case where the associative
commutative algebra $\fa_0$ contains the unit element $\e$,
and show the Lie antialgebra
$\fa$ has essentially no non-trivial central extensions in this case.

\begin{theorem}
\label{ZenT}
If the Lie antialgebra $\fa$ contains the unit element $\e\in\fa_0$,
then $\fa$ is a direct sum:
\begin{equation}
\label{splita}
\fa={\overline\fa}
\oplus{}Z(\fa),
\end{equation}
of its center and the Lie antialgebra $\overline\fa=\fa/Z(\fa)$
that has no non-trivial central extensions.
\end{theorem}

\begin{proof}
Let us consider the action of $\e$ on $\fa_1$.
The identity (\ref{CacT}) implies the ``half-projector''
relation:
$$
\textstyle
\ad^1_\e\circ\ad^1_\e=\half\,\ad^1_\e.
$$
Therefore, $\fa_1$ is split
$
\textstyle
\fa_1=\fa_{1,\half}\oplus\fa_{1,0}
$
to a sum of $\half$- and
$0$-eigenspaces, respectively.
That is, $\ad_e|_{\fa_{1,\half}}=\half\Id$ and
$\ad_e|_{\fa_{1,0}}=0$.

\begin{lemma}
\label{ZenTL}
The space $\fa_{1,0}$ coincides with the center of $\fa$.
\end{lemma}

\begin{proof}
Let $a\in\fa_{1,0}$, that is, $\e\cdot{}a=0$.
One has to show that $x\cdot{}a=0$ for all $x\in\fa$.

Let first $x=\a$ be an element of $\fa_0$.
The identity (\ref{CacT}) implies
$$
\e\cdot{}\left(\a\cdot{}a\right)=
\e\cdot{}\left(\a\cdot{}a\right)+
\a\cdot{}\left(\e\cdot{}a\right)=
\left(e\cdot{}\a\right)\cdot{}a=
\a\cdot{}a.
$$
But then $\ad^1_\e\circ\ad^1_\e=\half\,\ad^1_\e$ implies
$\a\cdot{}a=0$.

In the case where $x=b$ is an element of $\fa_0$ the proof is similar.
\end{proof}

Let us show now that
$\overline\fa$ has no non-trivial central extensions.

Let $C$ be a 2-cocycle of type I on $\fa$.
Apply the first identity (\ref{ExtI}) to $\b=\e$,
where $\e$ is the unit. One has
$$
C(\a,a)-C(\a,\e\cdot{}a)=
C(\e,\a\cdot{}a).
$$
If $a\in{}Z(\fa)$, this formula implies $C(\a,a)=0$.
If $a$ is an element of the $\half$-eigenspace $\fa_{1,\half}$ of the
unit element $\e$, then one obtains
$
C(\a,a)=
2\,C(\e,\a\cdot{}a).
$
Therefore, the cocycle $C$ is a coboundary.

Let now $C$ be a 2-cocycle of type II.
It can be decomposed into a pair $(C_0,C_1)$ of
maps
$$
C_0:\fa_0\otimes\fa_0\to{}\bbK,
\qquad
C_1:\fa_1\otimes\fa_1\to{}\bbK,
$$
where $C_0$ is symmetric and $C_1$ is skew-symmetric.
The first condition (\ref{ExtII}) gives
$$
C_0(\e,a\cdot{}b[)=
C_1(\e\cdot{}a,b)+C_1(a,\e\cdot{}b),
$$
so that $C_1(a,b)=C_0(\e,]a,b[)$.
The second condition (\ref{ExtII}) implies
$
C_0(\a,\b)=C_0(\e,\a\cdot{}\b).
$
Therefore, the cocycle $C$ is, again, a coboundary.

Proof of Theorem \ref{ZenT} is complete.
\end{proof}

\section{Classification results}\label{ClRS}

In this section, we prove that  $K_3(\bbC)$ is
the only finite-dimensional complex simple Lie
antialgebra.
In the real case one has $K_3(\bbR)$ and $K_3(\bbC)$.
This means that the situation is similar to the case of commutative
algebras.
We also prove that $K_3$ is characterized by the fact that its algebra of derivations
is isomorphic to $\osp(1|2)$.

We also obtain a complete classification of finite-dimensional Lie
antialgebras of rank~1, i.e., with
$\dim\fa_0=1$. 
This, in particular, provides with a number of examples
of Lie antialgebras, other that we already considered.

\subsection{Classification of finite-dimensional simple Lie
antialgebras}\label{UnSSe}

We call a Lie antialgebra $\fa$ \textit{simple} if it contains no ideal
except for the trivial one and $\fa$ itself.
The classification of finite-dimensional simple Lie
antialgebras is similar to that of commutative algebras.

\begin{theorem}
\label{SATm}
(i)
There exists a unique finite-dimensional complex simple Lie antialgebra.

(ii)
There are two finite-dimensional simple Lie antialgebras over $\bbR$.
\end{theorem}

\noindent
Recall that the only simple finite-dimensional commutative algebras
over $\bbR$ are $\bbR$ and $\bbC=\bbR+i\bbR$.
In the complex case there is only $\bbC$ itself.

Let us start with the complex case.
Let $\fa$ be a simple finite-dimensional Lie antialgebra.
We will first assume that the commutative algebra
$\fa_0$ has no nilpotent elements.
As it is very well known,
$\fa_0$ is of the form
$$
\fa_0=
\underbrace{\bbC\oplus\cdots\oplus\bbC}_{r\;\hbox{times}}.
$$
We will prove that if $\fa$ is simple, then $r=1$.

Choose a basis $\{\e_1,\ldots,\e_r\}$ in $\fa_0$ such that
$\e_i\cdot{}\e_j=\d_{ij}$.
As in Section \ref{LiePoisson}, one associates with each element $\e_i$ a
presymplectic form $\om_i$ on $\fa_1$ (see formula \ref{OmFor}).

The following statement shows that, if $r>1$, then the algebra $\fa$
contains a non-trivial ideal.
Consider the subspace $\ker\om_1\subset\fa_1$ consisting of the elements
$a\in\fa_1$ such that, for all $b$ one has: $a\cdot{}b$ is a combination of
$\e_i$ with $i\geq2$.
Consider also the following subspace of $\fa$:
$$
\cI=
\ker\om_1\oplus
\langle\e_2,\cdots,\e_r\rangle.
$$

\begin{lemma}
\label{ILm}
The subspace $\cI$ is an ideal of $\fa$.
\end{lemma}

\begin{proof}
One has to prove that

(a) $\a\cdot{}\cI\subset\cI$,
for an arbitrary $\a\in\fa_0$;

(b) $a\cdot{}\cI\subset\cI$, 
for an arbitrary $a\in\fa_1$.

Part (a).
Let $a\in\ker\om_1$, by identity (\ref{ICommT}), one has for an
arbitrary $b\in\fa_1$:
$$
\left(\a\cdot{}a\right)\cdot{}b
=
\a\cdot{}
\left(a\cdot{}b
\right)-
a\cdot{}\left(
\a\cdot{}b
\right).
$$
The both terms in the right-hand-side are combinations of $\e_i$ with
$i\geq2$. Therefore, one obtains $\a\cdot{}a\in\ker\om_1$.

Part (b).
Let $\a\in\langle\e_2,\cdots,\e_r\rangle$, then $\e_1\cdot{}\a=0$.
Let $a\in\fa_1$ be arbitrary,
one has to prove that, again, $\a\cdot{}a\in\ker\om_1$.
Choose an arbitrary $b\in\fa_1$ and consider again the
expression
$\left(\a\cdot{}a\right)\cdot{}b$.
Since this is an even element of $\fa$, it can be written in the form:
$$
\left(\a\cdot{}a\right)\cdot{}b=
\sum_{i=1}^r
c_i\,\e_i.
$$
One has to prove that $c_1=0$.
One has
$
\e_1\cdot{}
\left(
\left(\a\cdot{}a\right)\cdot{}b
\right)=
\left(\e_1\cdot{}
\left(\a\cdot{}a\right)\right)\cdot{}b+
\left(
\a\cdot{}a\right)\cdot{}
\left(
\e_1\cdot{}b
\right).
$
But the first summand in the right-hand-side vanishes.
Indeed, by (\ref{CacT}), one has
$$
\textstyle
\e_1\cdot{}
\left(
\a\cdot{}a\right)
=
\half
\left(
\e_1\cdot{}\a
\right)\cdot{}a
=0,
$$
since $\e_1\cdot{}\a=0$.
In the same way, one obtains
$$
\e_1\cdot{}
\left(
\e_1\cdot{}
\left(
\left(
\a\cdot{}a\right)\cdot{}
b
\right)\right)=
\left(
\a\cdot{}a\right)\cdot{}
\left(
\e_1\cdot{}
\left(
\e_1\cdot{}b
\right)\right).
$$
However, for the left-hand-side, one obtains using (\ref{AssCommT}):
$$
\e_1\cdot{}
\left(
\e_1\cdot{}
\left(
\left(
\a\cdot{}a\right)\cdot{}
b
\right)\right)=
\left(
\e_1\cdot{}
\e_1
\right)\cdot{}
\left(
\left(
\a\cdot{}a\right)\cdot{}
b
\right)=
\e_1\cdot{}
\left(
\left(
\a\cdot{}a\right)\cdot{}
b
\right)=
c_1
$$
since $\e_1\cdot{}\e_1=\e_1$;
while, for the right-hand-side, one gets using (\ref{CacT}):
$$
\textstyle
\left(
\a\cdot{}a\right)\cdot{}
\left(
\e_1\cdot{}
\left(
\e_1\cdot{}b
\right)\right)=
\half
\left(
\a\cdot{}a\right)\cdot{}
\left(
\left(
\e_1\cdot{}
\e_1\right)\cdot{}
b
\right)=
\half
\left(
\a\cdot{}a\right)\cdot{}
\left(
\e_1\cdot{}b
\right)=
\e_1\cdot{}
\left(
\left(
\a\cdot{}a\right)\cdot{}
b
\right)=
\half\,c_1.
$$
Therefore, $c_1=0$ and so $\a\cdot{}a\in\ker\om_1$.
The result follows.
\end{proof}

Lemma \ref{ILm} implies that $\dim(\fa_0)=1$.
Choose a non-zero element $\e\in\fa_0$ and denote $\om$ the corresponding 2-form. 

\begin{lemma}
\label{RLm}
The form $\om$ is of rank 2.
\end{lemma}

\begin{proof}
Choose a canonical (Darboux) basis $\{a_1\ldots,a_n,b_1,\ldots,b_n,c_1,\ldots,c_m\}$,
so that
$$
\om(a_k,b_\ell)=\d_{k\ell},
\qquad
\om(a_k,a_\ell)=\om(b_k,b_\ell)=\om(c_k,.)=0.
$$
where $k,\ell=1,\ldots,n$.
Let us show that $n=1$.
First, $n\geq1$, since otherwise $\om=0$ and $\fa_1$ is an ideal.
It follows that there are some elements $a_k$ and $b_k$ such that $\e=a_k\cdot{}b_k$.

Assume that $n>1$.
The identity (\ref{Jack}) implies
$$
\e\cdot{}a_\ell=
\left(a_k\cdot{}b_k\right)\cdot{}a_\ell=
-\left(b_k\cdot{}a_\ell\right)\cdot{}a_k
-\left(a_\ell\cdot{}a_k\right)\cdot{}b_k
=0
$$
for any $k\not=\ell$.
It follows that $\e\cdot{}a=0$, for any $a\in\fa_1$.
Furthermore,
$$
\e\cdot{}\e
=
\e\cdot{}
\left(
a_k\cdot{}b_k
\right)
=
\left(
\e\cdot{}
a_k
\right)\cdot{}b_k
+
a_k\cdot{}
\left(
\e\cdot{}
b_k
\right)=0,
$$
for any $i=1,\ldots,n$.
Therefore, $\e$ belongs to the center of $\fa$;
in particular, $\fa$ cannot be simple.
This is a contradiction.
\end{proof}

Lemma \ref{ILm} and Lemma \ref{RLm} imply Theorem \ref{SATm}
in the complex case, where the commutative algebra
$\fa_0$ has no nilpotent elements.
If now $\fa_0=\bbC^n\ltimes\cN$, where $\cN$ is a nilpotent ideal, then
the same arguments prove that $n\leq1$
and $\ker\om_1\oplus\cN$ is an ideal.
Theorem \ref{SATm} is proved
in the complex case.

The real case immediately follows from the complex one.
Indeed, let $\fa$ be a real simple Lie antialgebra,
the standard arguments show that the complexification
$\fa\otimes_\bbR\bbC$ is either simple or the direct sum of
two isomorphic simple ideals.

Theorem \ref{SATm} is proved.

\paragraph*{Proof of Theorem \ref{AsLThm}}

Let $\fa$ be a finite-dimensional commutative $\bbZ_2$-graded algebra
such that $\Der(\fa)\cong\osp(1|2)$.
As an $\osp(1|2)$-module, $\fa$ is a direct sum of irreducible modules.
Recall that finite-dimensional irreducible $\osp(1|2)$-modules
are the modules $\cD(h/2)$ with positive integer $h$.
These modules are of dimension $2h+1$ and of highest weight $h$.
It follows that for two elements of $\fa$ such that
$x_1\in\cD(h_1/2)$ and $x_2\in\cD(h_2/2)$,
one has:
\begin{equation}
\label{WeiPr}
x_1\cdot{}x_2\in\cD((h_1+h_2)/2).
\end{equation}
Since $\fa$ is finite-dimensional, there exists a non-zero submodule
$\cD(h'/2)\subset\fa$ with maximal~$h'$.

If $h'\not=0$, then (\ref{WeiPr}) implies that $\cD(h'/2)$
 belongs to the center of $Z(\fa)$.
 However, the Lie superalgebra $\End(Z(\fa))$ is a subalgebra
 of $\Der(\fa)$ so that $\Der(\fa)$ cannot be isomorphic to
 $\osp(1|2)$ if $\fa$ has a non-trivial center.

Theorem \ref{AsLThm} is proved.

\subsection{Lie antialgebras of rank 1}\label{Class}

Let us assume that the commutative algebra $\fa_0$ is one-dimensional.
There are two different possibilities:

\begin{enumerate}
\item
$
\fa_1=\bbK,
$
as a commutative algebra, so that it contains
the unit element denoted by $\e$, such that $\e\cdot\e=\e$;

\item
$\fa_1$ is nilpotent and the only odd generator $\a$
satisfies $\a\cdot\a=0$.

\end{enumerate}

The structure of the algebra $\fa$ is
characterized by one bilinear skew-symmetric form on $\fa_1$
defined by
$$
a\cdot{}b=
\om(a,b)\,\a,
$$
where $\a$ is a (unique up to a constant) non-zero element of $\fa_1$.

Let us first construct several examples of Lie antialgebras of rank 1.

\paragraph*{A. The form $\om$ is non-degenerate}

\textbf{(A1)}
The $(2n)|1$-dimensional nilpotent Lie
antialgebra with basis
$\{a_1,b_1,\ldots,a_n,b_n;\a\}$ that appeared in
the case a) of the above proof is characterized by the
relations
\begin{equation}
\label{HeiEq}
a_i\cdot{}b_j=\d_{ij}\,\a,
\end{equation}
where $i,j=1,\ldots,n$
and all other products of the basic elements vanish.
We call it the \textit{Heisenberg antialgebra}
and denote it by $\ah_n$.
This algebra is ample.

\begin{rmk}
{\rm
Notice, that the relations (\ref{HeiEq}) are exactly as those
of the standard Heisenberg Lie algebra, but the central element
$\a$ is odd.
As in the usual Lie case, the Heisenberg antialgebra $\ah_n$ is a
central
extension of type II of the abelian Lie antialgebra $\bbK^{2n}$.
}
\end{rmk}

\textbf{(A2)}
Another interesting example is a family of Lie antialgebras of
dimension $1|2$.
The basis of these Lie antialgebras will be denoted by $\{\a;\,a,b\}$;
the commutation relations are
\begin{equation}
\label{Redis}
a\cdot{}b=\a,
\qquad
\a\cdot{}a=\kappa\,b
\end{equation}
where $\kappa$ is a constant
and the other products vanish.
If $\kappa=0$, then this is just the Heisenberg antialgebra $\ah_1$,
if $\kappa\not=0$, then this is a non-trivial
deformation of $\ah_1$.

In the case $\bbK=\bbC$,
all of the Lie antialgebras  (\ref{Redis}) with $\kappa\not=0$
are isomorphic to each other.
We call this algebra \textit{twisted Heisenberg antialgebra}
and denote it by $\widetilde{\ah}_1$.

If $\bbK=\bbR$, however, the sign of $\kappa$ is an invariant.
We will assume:
$$
\kappa=1,
\quad
\hbox{if}
\quad
\bbK=\bbC,
\qquad
\kappa=\pm1,
\quad
\hbox{if}
\quad
\bbK=\bbR.
$$
One thus gets two different Lie antialgebras:
$\widetilde{\ah}^+_1$ and $\widetilde{\ah}^-_1$.

We are ready to formulate a partial result.

\begin{proposition}
\label{promezh}
The complete list of the real Lie antialgebras of rank 1 with a
non-degenerate 2-form $\om$ is as follows:
\begin{equation}
\label{ListA}
K_3,
\qquad
\ah_n,
\qquad
\widetilde{\ah}^+_1,
\qquad\widetilde{\ah}^-_1;
\end{equation}
in the complex case, the Lie antialgebras
$\widetilde{\ah}^+_1$ and $\widetilde{\ah}^-_1$ are isomorphic.
\end{proposition}

\begin{proof}
Consider first the case where $\fa_1$ is not nilpotent, i.e.,
$\a\cdot{}\a\not=0$.
We already proved that, in this case, $\fa$
is of dimension $1|2$, see Lemma~\ref{RLm}.
Therefore, $\fa=K_3$.

Assume that $\a\cdot{}\a=0$.
If $\om$ is of rank $n>1$, one proves,
in the same way as in Lemma~\ref{RLm}, that $\a\cdot{}a=0$ for all
$a$, so that $\fa=\ah_n$.

If, finally, $\om$ is of rank $1$, then the identity (\ref{CacT}) implies that
$(\ad^1_\a)^2=0$.
One then easily shows that any such operator
is equivalent to $\ad^1_\a$ in (\ref{Redis}) up to the area preserving
changes of the basis.
It follows that $\fa=\widetilde{\ah}^+_1$ or $\widetilde{\ah}^-_1$.
\end{proof}

\paragraph*
{B. The 2-form $\om$
is identically zero}

The Lie antialgebra $\fa$ is then
non-ample and determined by the operator $\ad_\a$.

\textbf{(B1)}
If $\fa_1$ contains the unit element $\e$, then $\fa$ is split
into a
direct sum (\ref{splita}).
The centerless Lie antialgebra $\overline{a}$ has the basis
$\{\e;\,a_1,\ldots,a_n\}$ with the following set of relations:
$$
\textstyle
\e\cdot{}\e=\e,
\qquad
\e\cdot{}a_i=\half\,a_i,
\qquad
a_i\cdot{}a_j=0.
$$
We call this Lie antialgebra the \textit{affine antialgebra} and denote
by
$\aaf(n)$. One then has
$
\fa=\aaf(n)\oplus{}Z,
$
where $Z$ is the 
center of
$\fa$.

\textbf{(B2)}
If $\a\cdot{}\a=0$ for all $\a\in\fa_0$, then $\ad^1_\a\circ\ad^1_\a=0$.
These are very degenerated Lie antialgebras and
their classification is equivalent to the classification
of nilpotent (of order 2) linear operators.
We do not discuss here this problem of linear algebra.

Let us summarize the above considerations.

\begin{proposition}
\label{promezhB}
A Lie antialgebras of rank 1 with $\om=0$ is one of
the following two classes:
\begin{equation}
\label{ListB}
\fa=\aaf(n)\oplus{}Z,
\qquad
\fa
\;
\hbox{is of type (B2)}.
\end{equation}
\end{proposition}

\paragraph*{C. The ``mixed case'' $2<\mathrm{rk}\,\om<\dim\fa_0$}

\textbf{(C1)}
Define a $1|3$-dimensional Lie antialgebra with basis
$\{\a;\,a,b,z\}$ and the relations
\begin{equation}
\label{ExAl1}
a\cdot{}b
=\a,
\qquad
\a\cdot{}a=z
\end{equation}
and all other products vanish.
We denote this Lie antialgebra $\widehat{\ah}_1$.
The element $z$ spans the center, so that this is a central extension
(of type I) of $\ah_1$.

\textbf{(C2)}
Define a $1|4$-dimensional Lie antialgebra
with the basis $\{\a;\,a,b,z_1,z_2\}$ and the relations
\begin{equation}
\label{ExAl2}
a\cdot{}b
=\a,
\qquad
\a\cdot{}a=z_1,
\qquad
\a\cdot{}b=z_2
\end{equation}
and all other products vanish.
We denote this Lie antialgebra $\widehat{\widehat{\ah}}_1$.
This is a central extension (of type I) of the above
algebra $\widehat{\ah}_1$.

We are now ready to formulate the main statement of this section.

\begin{theorem}
\label{ClThm}
A Lie antialgebra of rank 1 is of the form
$
\fa=\overline{\fa}\oplus{}Z,
$
where $\overline\fa$ belongs either to the list (\ref{ListA}), or to
the list (\ref{ListB}), or one of the antialgebras
$\widehat{\ah}_1,\widehat{\widehat{\ah}}_1$.
\end{theorem}

\begin{proof}
We already proved the theorem in the following two cases:
the form $\om$ is non-degenerate, or $\om\equiv0$.
It remains to consider the intermediate case where
the 2-form $\om$ is not
identically zero but with a non-trivial kernel:
$
\cI=\ker\om\not=\{0\}.
$
The space $\cI$ is then an abelian ideal (see Proposition
\ref{KerLem})
and, furthermore, belongs to the center (see Proposition \ref{ZenLem}).
We summarize this in a form of a

\begin{lemma}
\label{CeL}
The Lie antialgebra $\fa$ is a central extension of $\fa/\cI$.
\end{lemma}

To complete the classification, one now has to classify
the central extensions of type I of the antialgebras with
non-degenerate
form
$\om$, that is, of (\ref{HeiEq}) and (\ref{Redis}).

\begin{lemma}
\label{TrExLem}
The Lie antialgebras 
$\ah_n$ with $n\geq2$ and
$\widetilde{\ah}_1$
(resp. $\widetilde{\ah}^+_1,\widetilde{\ah}^-_1$)
have no non-trivial central extensions of type I.
\end{lemma}

\noindent
\begin{proof}
Let $C$ be a 2-cocycle of type I on $\ah_n$ with $n\geq2$.
One has
$$
C\left(\a,\,a_i\right)=
C\left(a_j\cdot{}b_j,\,a_i\right)=
-C\left(b_j\cdot{}a_i,\,a_j\right)
-C\left(a_i\cdot{}a_j,\,b_j\right)=0
$$
(from the second identity (\ref{ExtI})) for all $i\neq{}j$.
Similarly, $C(\a,b_i)=0$ for all $i$.
Therefore, $C$ is identically zero.

For the Lie antialgebra $\widetilde{\ah}_1$ and an arbitrary
2-cocycle $C$ of type I, one has
$$
\textstyle
C\left(\a,b\right)=
\frac{1}{\kappa}\,C\left(\a,\,\a\cdot{}a\right)=
\frac{1}{2\kappa}\,C\left(\a\cdot{}\a,\,a\right)=0.
$$
The cocycle $C$ is then defined by its values on $\a$ and $a$.
Let us show that the corresponding extension is trivial.
Let $C(\a,a)=cz$, where $z$ is an arbitrary generator of the center
and $c$ arbitrary constant.
Set $b'=b+\frac{c}{\kappa}\,z$. 
In the basis $\{\a;\,a,b',z\}$,
the cocycle $C'$ vanishes.
\end{proof}

\begin{lemma}
\label{HeiExtLem}
The algebra $\ah_1$ has a unique non-trivial central
extension of type I.
\end{lemma}
\begin{proof}
Indeed, let $C$ be a 2-cocycle of type I on $\ah_1$.
It is given by the formula
$$
C(\a,a)=c_1\,z,
\qquad
C(\a,b)=c_2\,z,
$$
where $z$ is an element of the center and $c_1$ and $c_2$
are arbitrary constants.
If $c_1\not=0$, then choose
another element of the basis: $b'=b+\frac{c_2}{c_1}\,a$.
One obtains $C(\a,b)=0$.
Now taking $z'=c_1z$, one gets precisely the Lie antialgebra~$\widehat{\ah}_1$.
This Lie antialgebra is not isomorphic to $\ah_1$ so that the extension
is, indeed, non-trivial.
\end{proof}

\begin{lemma}
\label{HeiExtBisLem}
The algebra  $\widehat{\widehat{\ah}}_1$ is the unique non-trivial central
extension of type I of $\widehat{\ah}_1$.
\end{lemma}
\begin{proof}
It is similar to the proof of Lemma \ref{HeiExtLem}.
\end{proof}

In the same way, one proves that the Lie antialgebra 
$\widehat{\widehat{\ah}}_1$ has no non-trivial central extensions
of type  I.
We thus classified all the non-trivial central extensions of
the Lie antialgebras of rank 1 with non-degenerate form $\om$.

Theorem \ref{ClThm} is proved.
\end{proof}

\vskip 0.5cm

\noindent \textbf{Acknowledgments}.
I am pleased to thank K. Bering, F. Chapoton, C. Conley, C. Duval, A. Elduque, Y. Fregier,
D. Fuchs, H. Gargoubi,  J. Germoni, K. Iohara, O. Kravchenko, 
M.~Kreusch, P.~Lecomte, S. Leidwanger,
D. Leites, S. Morier-Genoud,
S. Parmentier, N. Poncin, C.~Roger, S. Tabachnikov, A.~Tchoudjem, A. Vaintrob
and T. Voronov for enlightening discussions.


CNRS,
Institut Camille Jordan,
Universit\'e Lyon 1,

Villeurbanne Cedex, F-69622,
FRANCE;

ovsienko@math.univ-lyon1.fr

\end{document}